\documentclass[pra,twocolumn,amsmath,amssymb,showpacs,nofootinbib,floatfix]{revtex4}

\usepackage{graphicx,bm}

\makeatletter
\def\graphicscale{\twocolumn@sw{0.3}{0.4}}
\def\graphicthreescale{\twocolumn@sw{0.3}{0.4}}

\begin{document}

\title{Dimensional crossover of Bose-Einstein condensation phenomena \\

in quantum gases confined within slab geometries}

\author{Francesco Delfino and Ettore Vicari} 

\address{Dipartimento di Fisica dell'Universit\`a di Pisa
        and INFN, Largo Pontecorvo 3, I-56127 Pisa, Italy}

\date{\today}

\begin{abstract}

We investigate systems of interacting bosonic particles confined
within slab-like boxes of size $L^2\times Z$ with $Z\ll L$, at their
three-dimensional (3D) BEC transition temperature $T_c$, and below
$T_c$ where they experience a quasi-2D Berezinskii-Kosterlitz-Thouless
transition (at $T_{\rm BKT}<T_c$ depending on the thickness $Z$). The
low-temperature phase below $T_{\rm BKT}$ shows quasi-long-range
order: the planar correlations decay algebraically as predicted by the
2D spin-wave theory.  This dimensional crossover, from a 3D behavior
for $T\gtrsim T_c$ to a quasi-2D critical behavior for $T\lesssim
T_{\rm BKT}$, can be described by a transverse finite-size scaling
limit in slab geometries.  We also extend the discussion to the
off-equilibrium behavior arising from slow time variations of the
temperature across the BEC transition.  Numerical evidence of the
3D$\to$2D dimensional crossover is presented for the Bose-Hubbard
model defined in anisotropic $L^2\times Z$ lattices with $Z\ll L$.

\end{abstract}







\maketitle


\section{Introduction}
\label{intro}

The Bose-Einstein condensation (BEC) characterizes the low-temperature
behavior of three-dimensional (3D) bosonic gases, below a
finite-temperature BEC phase transition separating the
high-temperature normal phase and the low-temperature superfluid BEC
phase. The phase coherence properties of the BEC phase have been
observed by several experiments, see e.g.
Refs.~\cite{CWK-02,Andrews-etal-97,Stenger-etal-99,Hagley-etal-99,BHE-00,
  Dettmer-etal-01,Hellweg-etal-02,Hellweg-etal-03,Ritter-etal-07,BDZ-08}.
Several theoretical ad experimetal studies have also investigated the
critical properties at the BEC transition, when the condensate begins
forming, see, e.g.,
Refs.~\cite{DRBOKE-07,DZZH-07,BB-09,CV-09,ZKKT-09,
  Trotzky-etal-10,HZ-10,PPS-10,NNCS-10,ZKKT-10,QSS-10,FCMCW-11,HM-11,
  Pollet-12,CR-12,CTV-13,CN-14,CCBDWNDB-14,CNPV-15,
  NGSH-15,CCBDWNBD-15,DV-17,BN-17}. Both the phase-coherence
properties of the BEC phase and the critical behavior at the BEC
transition turn out to be particularly sensitive to the inhomogeneous
conditions arising from spatially-dependent confining potentials,
and/or the geometry of the atomic-gas system.  Inhomogeneous
conditions due to space-dependent trapping potentials give rise to a
universal distortion of the homogeneous critical behavior, which can
be cast in terms of a universal trap-size scaling~\cite{CV-09,CTV-13}
controlled by the same universality class of the 3D BEC transition.
In the case of homogeneous traps, such as those experimentally
realized in Refs.~\cite{CCBDWNDB-14,NGSH-15,CCBDWNBD-15,BN-17}, the
geometry of the trap may lead to quite different phase-coherence
properties, when passing from 3D, to quasi-2D, or quasi-1D systems.
For example, atomic gases in elongated homogeneous
boxes~\cite{CDMV-16} and harmonic
traps~\cite{PSW-01,Dettmer-etal-01,Hellweg-etal-02,Mathey-etal-10,
  GCP-12,RMHDTLK-13} show a dimensional crossover from a
high-temperature 3D behavior to a low-temperature quasi-1D behavior.

In this paper we consider bosonic particle systems confined within
slab geometries, i.e. within boxes of size $L^2\times Z$ with $Z\ll
L$.  We investigate their behavior at the BEC transition temperature
$T_c$ (this is the critical temperature of the 3D system in the
thermodynamic limit, i.e. when all system sizes tend to infinity) and
at lower temperatures. Their low-temperature behavior ($T<T_c$) is
further characterized by the possibility of undergoing a
finite-temperature transition to a quasi-long range order (QLRO)
phase, with long-range planar correlations which decay algebraically.
This is the well-known Berezinskii-Kosterlitz-Thouless (BKT)
transition~\cite{KT-73,B-72,Kosterlitz-74,JKKN-77}, which occurs in 2D
statistical systems with a global U(1) symmetry.  Experimental
evidences of BKT transitions have been also reported for quasi-2D
trapped atomic
gases~\cite{HKCBD-06,KHD-07,HKCRD-08,CRRHP-09,HZGC-10,Pl-etal-11,Desb-etal-12}.

The behavior of homogeneous gases in slab geometries can be described
in terms of a dimensional crossover, from 3D behaviors for $T\gtrsim
T_c$ to a quasi-2D critical behavior for $T\lesssim T_{\rm BKT}$.  In
the limit of large thickness $Z$, the quasi-2D BKT transition
temperature approaches that of the 3D BEC transition, i.e.  $T_{\rm
  BKT}\to T_c$ for $Z\to \infty$ (assuming the thermodynamic limit for
the planar directions, i.e. $L\gg Z$).  The interplay of the BEC and
BKT critical modes gives rise to a quite complex behavior. We show
that this can be described by a transverse finite-size scaling (TFSS)
limit for systems in slab geometries~\cite{Barber-83,Privman-90},
i.e., $Z\to\infty$ and $T\to T_c$ keeping the product
$(T-T_c)Z^{1/\nu}$ fixed, where $\nu$ is the correlation-length
exponent at the 3D BEC transition.  In this TFSS limit the BKT
transition below $T_c$ appears as an essential singularity of the TFSS
functions.  The dimensional-crossover scenario is expected to apply to
any quantum gas of interacting bosonic particles confined in boxes or
lattice structures with slab geometries.  Analogous arguments apply to
$^4$He systems in film geometries~\cite{GKMD-08}, and to 3D $XY$ spin
models defined in lattices with slab
geometries~\cite{SM-95,SM-96,SM-97,Hasenbusch-09}.

We also extend the discussion to the off-equilibrium behavior arising
from slow time variations of the temperature across the BEC
transition.  The behavior of weakly interacting atomic gas
confined in quasi-2D geometries has been experimentally investigated
under time-dependent protocols across the BEC regime, see, e.g.,
Refs.~\cite{CCBDWNBD-15,BN-17}, to verify the Kibble-Zurek mechanism
of defect production~\cite{Kibble-76,Zurek-85}.  In gases confined
within slab geometries, the off-equilibrium behavior arising from the
slow variation of the temperature across the BEC transition point is
made particularly complex by the presence of the quasi-2D BKT
transition at $T_{\rm BKT}\lesssim T_c$.  Thus, disentangling the
behaviors corresponding to 3D BEC and quasi-2D BKT transitions may be
quite hard in experimental or numerical analyses.  To describe this
complex behavior, we put forward the emergence of an off-equilibrium
transverse finite-size scaling for bosonic gases confined
within slab-like homogeneous traps.

We provide evidence of the dimensional-crossover scenario in quantum
gases by a numerical study of the Bose-Hubbard (BH)
model~\cite{FWGF-89}, which models gases of bosonic atoms in optical
lattices~\cite{JBCGZ-98,BDZ-08}.  We show that the predictions of the
3D$\to$2D dimensional crossover are realized when considering anisotropic
slab-like lattices $L^2\times Z$ with $Z\ll L$.  With decreasing $T$
from the high-temperature normal phase, we first meet a quasi-BEC
transition where the critical length scale $\xi$ gets large, but it
does not diverge being limited by $\xi\sim Z$ (keeping $Z$ fixed).
Then we observe a BKT transition to a QLRO phase, where the system
develops planar critical correlations essentially described by a
Gaussian spin-wave theory.  The dimensional crossover
explains the apparently complex behavior of the one-particle
correlation functions and the corresponding length scale, when
decreasing the temperature from $T>T_c$, where $T_c$ is the 3D BEC
transition temperature, to $T < T_{\rm BKT}<T_c$, where $T_{\rm BKT}$
depends on the thickness $Z$. The results are also consistent with
the scaling predictions of the TFSS theory for systems in slab
geometries.

The paper is organized as follows.  In Sec.~\ref{moobs} we introduce
the BH model that we use as a paradigmatic model of Bose gases showing
the phenomenon of dimensional crossover in slab geometries.  In
Sec.~\ref{phdc} we present the general theory of the dimensional
crossover in slab geometries.  In Sec.~\ref{tssslab} we discuss the
new features arising from the presence of a harmonic trap along the
shorter transverse direction.
Sec.~\ref{lowtbh} reports some exact spin-wave results for the
phase-coherence correlations within the low-temperature phase of
quasi-2D systems with U(1) symmetry. In Sec.~\ref{KZ} we discuss the
off-equilibrium behavior arising from slow time variations of the
temperature across the BEC transition.  In Sec.~\ref{dimcross} we
provide numerical evidences of the dimensional crossover in 3D BH
models defined on lattices with slab geometries.  Finally, we
summarize our results in Sec.~\ref{conclu}.

\section{The Bose-Hubbard model in slab geometries}
\label{moobs}

Lattice BH models~\cite{FWGF-89} are interesting examples of
interacting Bose gases undergoing BEC transitions. They provide
realistic models of gases of bosonic atoms in optical
lattices~\cite{JBCGZ-98}.  In the following discussions we use the BH
model as a paradigmatic model of Bose gases showing the dimensional
crossover in slab geometries.

The Hamiltonian of BH models reads
\begin{eqnarray}
H_{\rm BH} &=& - t \sum_{\langle ij\rangle} (b_i^\dagger b_j+
b_j^\dagger b_i) + \label{bhm}\\
&+&{U\over 2} \sum_i n_i(n_i-1) - \mu \sum_i n_i\,,
\nonumber
\end{eqnarray}
where $b_i$ is a bosonic operator, $n_i\equiv b_i^\dagger b_i$ is the
particle density operator, the sums run over the bonds ${\langle ij
  \rangle }$ and the sites $i$ of a cubic $L_1\times L_2\times L_3$
lattice, $a=1$ is the lattice spacing.  
The phase coherence properties can be inferred from the one-particle
correlation function
\begin{eqnarray}
G({\bm r}_1,{\bm r}_2) \equiv {{\rm Tr}\; b_{{\bm r}_1}^\dagger b_{{\bm r}_2}
e^{-H_{\rm BH}/T}\over 
{\rm Tr}\; e^{-H_{\rm BH}/T}}.
\label{gbdef}
\end{eqnarray}
We set the hopping parameter $t=1$, so that all energies are expressed
in units of $t$, and the Planck constant $\hslash=1$.

\begin{figure}
\includegraphics{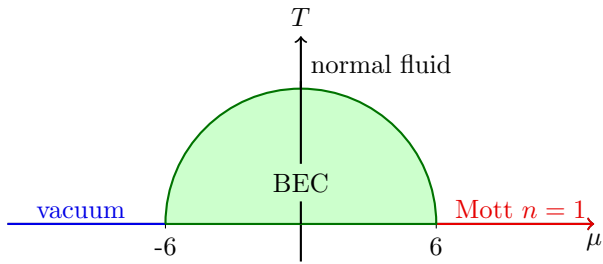}
\caption{Sketch of the $T$-$\mu$ (in unit of the hopping parameter
  $t$) phase diagram of the 3D BH model in the hard-core $U\to\infty$
  limit.  The BEC phase is restricted to a finite region between
  $\mu=-6$ and $\mu=6$.  It is bounded by a BEC transition line
  $T_c(\mu)$, which satisfies $T_c(\mu)=T_c(-\mu)$ due to a
  particle-hole symmetry.  Its maximum occurs at $\mu=0$,
  where~\cite{CN-14,CNPV-15} $T_c(\mu=0)= 2.01599(5)$; we also know
  that~\cite{CTV-13} $T_c(\mu\pm 4)=1.4820(2)$.  At $T=0$ two further
  quantum phases exist: the vacuum phase ($\mu<-6$) and the
  incompressible $n=1$ Mott phase ($\mu>6$).  }
\label{3dphasedia}
\end{figure}

The phase diagram of 3D BH models and their critical behaviors have
been much investigated, see
e.g. Refs.~\cite{FWGF-89,CPS-07,CR-12,CTV-13,CN-14,CNPV-15}.  Their
$T$-$\mu$ phase diagram presents a finite-temperature BEC transition
line. This is characterized by the accumulation of a macroscopic
number of atoms in a single quantum state, which gives rise to a
phase-coherent condensate.  See for example Fig.~\ref{3dphasedia},
which shows a sketch of the phase diagram of 3D BH models in the
hard-core $U\to\infty$ limit, where the occupation site number is
limited to the cases $n=0,\,1$.  The condensate wave function provides
the complex order parameter of the BEC transition, whose critical
behavior belongs to the U(1)-symmetric $XY$ universality class. This
implies that the length scale $\xi$ of the critical modes diverges at
$T_c$ as~\cite{PV-02,Lipa-etal-96,CHPV-06,BMPS-06,GZ-98,KPSV-16,KP-17}
\begin{equation}
\xi\sim (T-T_c)^{-\nu},\qquad \nu=0.6717(1).
\label{xi3d}
\end{equation}
This has been accurately verified by numerical studies, see, e.g.,
Refs.~\cite{CR-12,CTV-13,CN-14,CNPV-15}.  The BEC phase extends below
the BEC transition line. In particular, in the hard-core limit
$U\to\infty$ and for $\mu=0$ (corresponding to half filling), the BEC
transition occurs at~\cite{CN-14,CNPV-15} $T_c=2.01599(5)$.

We consider BH lattice gases in anisotropic slab-like geometries,
i.e. $L^2\times Z$ lattices with $Z \ll L$.  We consider open boundary
conditions (OBC) along the transverse $Z$-direction; we label the
corresponding coordinate as $-(Z-1)/2\le z \le (Z-1)/2$, so that the
innermost plane is the $z=0$ plane.  This choice is motivated by the
fact that OBC correspond to gas systems trapped by hard walls, such as
the experimental systems of
Refs.~\cite{CCBDWNDB-14,NGSH-15,CCBDWNBD-15,BN-17}.  Since the
thickness $Z$ of the slab is generally considered as much smaller than
the size $L$ of the planar directions, and in most cases we consider
the 2D {\em thermodynamic} $L\to\infty$ limit keeping $Z$ fixed, the
boundary conditions along the planar directions are generally
irrelevant for our study around $T_c$. However, they become relevant
at the BKT transition where the planar correlation length diverges.
In the following we consider the most convenient periodic boundary
conditions (PBC) along the large planar dimensions; the corresponding
site coordinates are ${\bm x}=(x_1,x_2)$ with $x_{1,2}=1,...,L$.

We want to understand how the phase diagram and critical behavior
change when varying the thickness $Z$.  As we shall argue, BH systems
below $T_c$ are expected to develop quasi-2D critical modes, leading
to a BKT transition with a diverging planar correlation length, and a
low-temperature QLRO phase. 

To study this phenomenon, and in particular how the $Z\to\infty$ limit
eventually realizes the 3D critical behavior at $T_c$, we focus on the
behavior of the correlation function (\ref{gbdef}) along the planar
directions.  In particular, for simplicity reasons, we study the
correlation function between points belonging to the central $z=0$
plane, i.e.
\begin{equation}
g({\bm x}_1-{\bm x}_2) \equiv G[({\bm x}_1,0), ({\bm x}_2,0)],
\label{gx1x2}
\end{equation}
where we have taken into account the invariance of the system for
translations along the $\hat{1}$ and $\hat{2}$ directions.  In
particular, we consider the {\em planar susceptibility}
\begin{equation}
\chi = \sum_{\bm x} g({\bm x})  
\label{chi}
\end{equation}
and the {\em planar second-moment correlation length} $\xi$ 
\begin{equation}
\xi^2 = {1\over 4\chi} \sum_{\bm x} {\bm x}^2 g({\bm x}).
\label{xidef}
\end{equation}
More precisely, since we consider PBC along the planar directions,
we use the equivalent definition
\begin{equation} \label{eq:xi_definition} 
 \xi^2 \equiv \frac{1}{4 \sin^2 (\pi/L)} 
      \frac{\tilde{g}({\bm 0}) - \tilde{g}({\bm p})} 
      {\tilde{g}({\bm p})}, 
\end{equation} 
where ${\tilde{g}(\bm{p})}$ is the Fourier transform of $g({\bm x})$, and
${\bm p}=(2\pi/L,0)$.

The helicity modulus $\Upsilon$ is a measure of the response of the
system to a phase-twisting field along one of the lattice
directions~\cite{FBJ-73}. In the case of bosonic systems, it is
related to the superfluid density~\cite{FBJ-73,PC-87,CDMV-16}.
We consider the helicity modulus
along the planar directions $\hat{1}$ and $\hat{2}$, i.e.,
\begin{eqnarray}
\Upsilon_a \equiv  \frac{1}{Z} \left. \frac{\partial^2
      F(\phi_a)}{\partial\phi_a^2} \right|_{\phi_a=0} \equiv  
{T\over Z} Y_a,
      \label{helma}
\end{eqnarray}
where $F=-T\ln Z$ is the free energy, $\phi_a$ are twist angles along
one of the planar directions. Note that $Y_1=Y_2$ by symmetry for
$L^2\times Z$ systems.

As we shall see, the quantities
\begin{equation}
Y\equiv Y_a,\qquad R_L\equiv \xi/L, 
\label{Rdef}
\end{equation}
are particularly useful to check the effective spin-wave behavior
along the planar directions for $T\le T_{\rm BKT}<T_c$.

\section{Dimensional crossover of Bose gases in slab geometries}
\label{phdc}

\subsection{Phase diagram  for a finite thickness $Z$}
\label{finthick}

The 3D scenario sketched in Fig.~\ref{3dphasedia} substantially
changes if we consider a quasi-2D thermodynamic limit,
i.e. $L\to\infty$ keeping $Z$ fixed.  Indeed the length scale $\xi$
remains finite at the BEC transition point when $Z$ is kept fixed.  Of
course the full 3D critical behavior must be somehow recovered when
$Z\to\infty$, for which one expects $\xi(Z) \sim Z$. More precisely,
defining
\begin{equation}
R_Z = {\rm lim}_{L\to\infty} \;\xi/Z,
\label{rzdef}
\end{equation} 
standard FSS arguments~\cite{Barber-83,Privman-90}
predict that at the 3D critical point $T_c$
\begin{equation}
R_Z(T_c)  = R_Z^* + O(Z^{-\omega})
\label{xiz}
\end{equation}
where $R_Z^*$ is a universal constant and $\omega=0.785(20)$ is the
scaling-correction exponent associated with the leading irrelevant
perturbation at the $XY$ fixed point~\cite{GZ-98,PV-02,CHPV-06}.  Note
that the universal constant $R_Z^*$ depends on the boundary conditions
along the transverse direction (the boundary conditions along the
planar directions are irrelevant since we assume $L\gg Z$ and $\xi\sim
Z$).

However, we should also take into account that 2D or quasi-2D systems
with a global U(1) symmetry may undergo a finite-temperature
transition described by the BKT
theory~\cite{KT-73,B-72,Kosterlitz-74,JKKN-77}.  The BKT transition
separates a high-temperature normal phase and a low-temperature phase
characterized by QLRO, where correlations decay algebraically at large
distances, without the emergence of a nonvanishing order
parameter~\cite{MW-66,H-67}.  When approaching the BKT transition
point $T_{\rm BKT}$ from the high-temperature normal phase, these
systems develop an exponentially divergent correlation length:
\begin{equation}
\xi \sim \exp\left( {c/\sqrt{\tau}}\right),\qquad 
\tau\equiv T/T_{\rm BKT}-1,
\label{xi2d}
\end{equation}
where $c$ is a nonuniversal constant. The magnetic susceptibility
diverges as $\chi\sim\xi^{7/4}$, corresponding to the critical
exponent $\eta=1/4$.  

\begin{figure}
\includegraphics*[scale=\graphicscale]{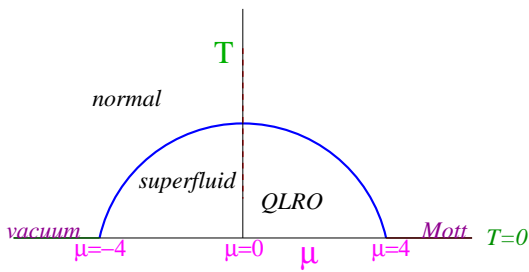}
\caption{ Sketch of the phase diagram of the 2D BH model in the
  hard-core $U\to\infty$ limit.  The normal and superfluid QLRO phases
  are separated by a finite-temperature BKT transition line, which
  satisfies $T_{\rm BKT}(\mu)=T_{\rm BKT}(-\mu)$ due to a
  particle-hole symmetry.  Its maximum occurs at $\mu=0$,
  where~\cite{CNPV-13} $T_{\rm BKT}(\mu=0)= 0.6877(2)$.  The
  superfluid QLRO phase is restricted to a finite region between
  $\mu=-4$ and $\mu=4$, which is narrower than that of the 3D phase
  diagram, see Fig.~\ref{3dphasedia}.  }
\label{2dphasedia}
\end{figure}

Consistently with the above picture, 2D BH systems [corresponding to
  the Hamiltonian (\ref{bhm}) with $Z=1$] undergo a BKT transition.
Fig.~\ref{2dphasedia} shows a sketch of the phase diagram of 2D BH
systems in the hard-core $U\to\infty$ limit.  The finite-temperature
BKT transition of BH models has been numerically investigated by
several studies, see e.g.
Refs.~\cite{CNPV-13,CR-12,HK-97,Ding-92,DM-90}.  In particular,
$T_{\rm BKT}=0.6877(2)$ in the hard-core $U\to\infty$ limit and for
$\mu=0$~\cite{CNPV-13}.  Note that the 2D BH systems do not show a
real BEC below the critical temperature $T_{\rm BKT}$, but QLRO where
the phase-coherence correlations decay algebraically.

The phase diagram of quasi-2D systems with finite thickness $Z>1$ is
expected to be analogous to that of 2D BH systems, with a BKT
transition at $T_{\rm BKT}$ depending on the thickness
$Z$.  Analogously to 2D systems, they are expected to show a QLRO
phase below $T_{\rm BKT}$, where correlation functions show power-law
decays along the planar directions, as described by the 2D spin-wave
theory.

\subsection{Dimensional crossover limit}
\label{fssslab}

The above scenario can be interpreted as a dimensional crossover from
a 3D behavior when $T\gtrsim T_c$, and $\xi$ is finite (in particular
the anisotropy of the system is not locally relevant when $\xi\ll Z$),
to an effective 2D critical behavior at $T\lesssim T_{\rm BKT}(Z)$
where the planar correlation length $\xi$ diverges.
  
Such a dimensional crossover can be described by an appropriate TFSS
limit, defined as $\delta\equiv 1 - T/T_c \to 0$ and $Z\to \infty$,
keeping $\delta Z^{1/\nu}$ fixed.  In this TFSS
limit~\cite{Barber-83,Privman-90}
\begin{equation}
R_Z\equiv \xi/Z \approx {\cal R}(X), \qquad X= Z^{1/\nu}\delta ,
\label{crosslimit}
\end{equation}
where ${\cal R}(X)$ is a universal function (apart from a trivial
normalization of the argument $X$), but depending on the boundary
conditions along the $Z$ direction.  Scaling corrections are
suppressed as $Z^{-\omega}$, analogously to Eq.~(\ref{xiz}).

In this TFSS framework the BKT transition appears as an essential
singularity of the scaling function ${\cal R}(X)$:
\begin{equation}
{\cal R}(X) \sim \exp\left({b\over \sqrt{X_{\rm BKT}-X}}\right)
\quad {\rm for}\;\;X\to X_{\rm BKT}^-,
\label{fxixs}
\end{equation}
where $X_{\rm BKT}$ is the value of the scaling variable
$X$ corresponding to the BKT transition point
\begin{eqnarray}
\delta_{\rm BKT}(Z) \equiv {T_c - T_{\rm BKT}(Z) \over T_c},
\label{deltabkt}
\end{eqnarray}
i.e.,
\begin{eqnarray}
X_{\rm BKT} = {\rm lim}_{Z\to\infty} Z^{1/\nu}\delta_{\rm BKT}(Z) \; > 0.
\label{xbkt}
\end{eqnarray}
The constant $b$ in Eq.~(\ref{fxixs}) is a nonuniversal constant
depending on the normalization of the scaling variable $X$.  ${\cal
  R}(X)$ is not defined for $X\ge X_{\rm BKT}$.  Note that the above
scaling equations predict that~\cite{Fisher-71,CF-76}
\begin{equation}
\delta_{\rm BKT}(Z) \sim Z^{-1/\nu}
\label{tctbkt}
\end{equation}
in the large-$Z$ limit.

The TFSS of the planar two-point function (\ref{gx1x2}) is given by
\begin{equation}
g({\bm x},Z)  \approx Z^{-(1+\eta)} {\cal G}({\bm x}/Z,X),
\label{gtfss}
\end{equation}
where $\eta=0.0381(2)$ is the critical exponent of the 3D $XY$ 
universality class~\cite{CHPV-06}, associated with the power-law decay
of the two-point function at $T_c$.  Eq.~(\ref{gtfss}) also implies
that the planar susceptibility defined as in Eq.~(\ref{chi}) behaves
as
\begin{equation}
\chi \approx Z^{1-\eta} f_\chi(X).
\label{chitfss}
\end{equation}

It is important to note that the above features are shared with any
quasi-2D statistical system with a global U(1) symmetry, and in
particular standard O(2)-symmetric spin models. Numerical analyses of
dimensional crossover issues for the $XY$ model are reported in
Refs.~\cite{SM-95,SM-96,SM-97,Hasenbusch-09}.

\section{Bose gases confined by a transverse harmonic trap}
\label{tssslab}

We now discuss the case of quasi-2D gases trapped by a harmonic
potential along the transverse direction, analogously to the
experimental setup of Ref.~\cite{CCBDWNBD-15}.

\subsection{The BH model in a transverse harmonic trap}
\label{BHtra}

In the case of the BH model the presence of a 
space-dependent trapping potential can be
taken into account by adding a further Hamiltonian term to
Eq.~(\ref{bhm}), i.e.
\begin{eqnarray}
H_{\rm hBH} = H_{\rm BH} + \sum_i V(z_i) n_i, \label{bhmt}
\end{eqnarray}
\begin{eqnarray}
V(z)= |z/\ell|^p,\label{potential}
\end{eqnarray}
where $z_i$ is the distance of the site $i$ from the central plane,
$p>0$, and $\ell$ can be considered as the transverse trap size.  The
harmonic potential corresponds to $p=2$.  The transverse trapping
potential coupled to the particle density turns out to be equivalent
to an effective chemical potential depending on the transverse
coordinate $z$,
\begin{equation}
\mu_e(\mu,z) \equiv  \mu - V(z).
\label{mue}
\end{equation}
Far from the central $z=0$ plane, the potential $V(z)$ diverges, thus
$\mu_e\to -\infty$ therefore $\langle n_i\rangle$ vanishes and the
particles are trapped along the transverse direction.  

We discuss the behavior of the system in the limit of infinite size of
the planar dimensions, along which the system appears as homogeneous.
For practical realizations, this regime may be realized by considering
hard-wall traps along the planar directions with size $L\gg \ell$
(more precisely $L\gg \ell^\theta$ where the exponent $\theta<1$ is
given below).

The planar correlation functions, for example along the $z=0$ plane,
are expected to behave similarly to the case of transverse hard-wall
traps.  With decreasing $T$ from the high-temperature normal phase,
the length scale $\xi$ gets large around the BEC transition
temperature $T_c$ (i.e. the critical temperature of the BEC transition
of the corresponding homogeneous 3D system). But it does not diverge,
since  $\xi\sim \ell^\theta$ where $\theta$ is an
appropriate exponent, see below. Then one may observe a BKT transition
to a QLRO phase around the $z=0$ plane, at $T_{\rm BKT}<T_c$ depending
on $\ell$.  In particular, in the extreme
$\ell\to 0$ limit, where all particles are confined within the $z=0$
plane, we recover the homogeneous 2D BH model, i.e. the model
(\ref{bhm}) with $Z=1$.  On the other hand, in the opposite
$\ell\to\infty$ limit, we again expect that $T_{\rm BKT}(\ell)\to
T_c^-$, analogously to the homogeneous case.  Therefore, similarly to
the homogeneous case, the system passes from a high-temperature 3D
behavior to a quasi-2D critical temperature at low temperature. This
change of regime may be also related to a transverse condensation
phenomenon~\cite{CCBDWNBD-15,BN-17,vDK-97,AJKB-11,RMHDTLK-13}.

\subsection{Transverse trap-size scaling}
\label{TTSS}

Like homegeneous systems with transverse hard-wall boundary
conditions, the critical behavior of the 3D system must be somehow
recovered in the large-$\ell$ limit, in a spatial region sufficiently
close to the central $z=0$ plane.  We argue that this limit can be
described by a universal transverse-trap-size scaling (TTSS), similar
to the TFSS limit discussed in Sec.~\ref{fssslab}.  To derive the TTSS
laws for the case at hand, we can exploit the same arguments 
used to derive
the trap-size scaling for isotropic
traps~\cite{CV-09,Pollet-12,CTV-13}.

The trapping potential (\ref{potential}) coupled to the particle
density significantly affects the critical modes, introducing another
length scale $\ell$.  Like general critical phenomena, see, e.g.,
Ref.~\cite{PV-02}, the asymptotic scaling behavior of the length scale
at $T_c$ is expected to be characterized by a power law:
\begin{equation}
\xi_t \sim \ell^\theta.
\label{xieltheta}
\end{equation}
The exponent $\theta$ can be determined by a scaling analysis of the
perturbation associated with the external potential coupled to the
particle density.  Its derivation is identical to that reported in
Refs.~\cite{CV-09,CTV-13} for isotropic traps.  The exponent $\theta$
turns out to be related to the correlation-length exponent $\nu$ of
the universality class of the critical behavior of the homogeneous BEC
transition, i.e.,
\begin{equation}
\theta = {p\nu\over 1 + p \nu},
\label{theta}
\end{equation}
where $\nu=0.6717(1)$ is the correlation-length exponent of the 3D $XY$
universality class. For harmonic transverse traps, i.e. $p=2$,
$\theta=0.57327(4)$.

On the basis of these TTSS arguments, we expect that the asymptotic
large-$\ell$ behavior of the two-point function around the central
$z=0$ plane, and in particular the correlation function defined as in
Eq.~(\ref{gx1x2}), behaves as
\begin{equation}
g({\bm x},\ell)  \approx \xi_t^{-(1+\eta)} 
{\cal G}_p({\bm x}/\xi_t,\delta \xi_t^{1/\nu}),
\label{gtrap}
\end{equation}
where $\xi_t\sim \ell^\theta$, $\delta \equiv 1 - T/T_c$, and we have
assumed that the planar sizes are infinite. Actually, one may also
take into account the planer size $L$ by adding a further scaling
variable $L/\ell^\theta$; the $L\to\infty$ scaling behavior
(\ref{gtrap}) is recovered when $L/\ell^\theta\gg 1$.  

The TTSS of the two-point function implies that the planar
second-moment correlation length along the $z=0$ plane, defined as in
eq.~(\ref{eq:xi_definition}), behaves asymptotically as
\begin{equation}
\xi_t \approx \ell^\theta {\cal R}_p({\cal X}),\qquad
{\cal X}\equiv \delta \ell^{\theta/\nu}.
\label{xiscatrap}
\end{equation}
In particular, we recover $\xi_t\sim \ell^\theta$ at $T_c$.  Note that
this scaling behavior is analogous to that of hard-wall traps,
cf. Eq.~(\ref{crosslimit}), with the transverse size $Z$ replaced by
$\ell^\theta$.  The leading corrections to the above asymptotic TTSS
are $O(\ell^{-\omega\theta})$. 

Note that the trap-exponent $\theta$ reported in Eq.~(\ref{theta}) is
identical to that of isotropic traps~\cite{CTV-13}, i.e. it does not
depend on the number of coordinates entering the space-dependence of
the inhomogeneous power-law potential coupled to the particle
density. However, the scaling functions ${\cal G}_p$ and ${\cal R}_p$,
entering Eqs.~(\ref{gtrap}) and (\ref{xiscatrap}), must definitely
differ.  Actually, in the $p\to\infty$ limit we must recover the TFSS
behavior, i.e. that of the homogeneous conditions along the transverse
direction with OBC, see Sec.~\ref{fssslab}. Since $\theta\to 1$ for
$p\to\infty$, $\ell \approx Z$ of the transverse hard-wall conditions.

The TTSS functions must present a singularity related to the BKT
transition for $T_{\rm BKT}<T_c$, unlike those of the isotropic TSS
because no such transition occurs for isotropic traps.  In particular,
TTSS implies that
\begin{equation}
\delta_{\rm BKT}(\ell) \equiv 1 - T_{\rm BKT}(\ell)/T_c \sim \ell^{-\theta/\nu},
\label{dbkt}
\end{equation}
and the TTSS function $f_\xi$ of (\ref{xiscatrap}) 
must show a BKT-like singularity at
\begin{equation}
{\cal X}_{\rm BKT}   = {\rm lim}_{\ell\to\infty}
\;\delta_{\rm BKT}(\ell) \,\ell^{\theta/\nu},
\label{bkttrasi}
\end{equation}
such as that reported in Eq.~(\ref{fxixs}).

\subsection{Criticality at the boundary of the BEC region}
\label{bousca}

Other interesting features arise at the boundary of the BEC region in
atomic gases confined by a transverse harmonic trap.  If the trap is
sufficiently large and the temperature is sufficiently low, different
phases may coexist in different space regions, when moving from the
central $z=0$ plane of the trap.  Indeed, due to the fact that the
effective chemical potential $\mu_e(z)$, cf. Eq.~(\ref{mue}),
decreases with increasing $z$, the BEC region is generally spatially
limited.  When moving from the $z=0$ plane, the quantum gas passes
from the BEC phase around the center of the trap (where space
coherence is essentially described by spin waves) to a normal phase
far from the center.  The atomic gas is expected to develop a peculiar
critical behavior at the boundary of the BEC region, with a nontrivial
scaling behavior controlled by the universality class of the
homogenous BEC transition in the presence of an effective linear
external potential coupled to the particle density~\cite{DV-17}.

This occurs around the planes where the distance $|z|$ from the $z=0$
plane is such that $T[\mu_e(\mu,z)]$ is equal to the BEC trasition
temperature at the local chemical potential $\mu_e(\mu,z)=\mu -
(z/\ell)^2$, i.e. when
\begin{equation}
T_c[\mu_e(\mu,z)] \approx T < T_c(\mu).
\label{tcmue}
\end{equation}
For example, consider the hard-core BH lattice gas (\ref{bhmt}) for
$\mu\le 0$ and $T<T_c(\mu)$, see Fig.~\ref{3dphasedia}.  Since
$T_c(\mu)$ decreases with decreasing $\mu$, a plane exists at distance
$z=z_b$ such that $T_c[\mu_e(\mu,z_b)] = T$, thus 
\begin{equation}
z_b= \ell \sqrt{\mu-\bar{\mu}},
\label{zbsol}
\end{equation}
 where $T_c(\bar{\mu})=T$. This plane separates
the BEC region from the normal-fluid region.  As argued in
Ref.~\cite{DV-17}, in the limit of large $\ell$,
the correlation functions around the surface where $T_c[\mu_e({\bm r})]=T$
are expected to develop a peculiar critical behavior in the presence of an
external effectively linear potential coupled to the particle density.

Around $z=z_b$
\begin{equation}
V(z) = V(z_b) + \Delta z/\ell_b + O[(\Delta z/\ell_b)^2]
\label{vzzb}
\end{equation}
with
\begin{equation}
\ell_b = {\ell\over 2 \sqrt{\mu-\bar{\mu}}}.
\label{Dez}
\end{equation}
The critical behavior at the critical planes $z=z_b$ is
essentially determined by the linear term
\begin{equation}
V_b = \Delta z/\ell_b, \qquad \Delta z \equiv z - z_b.
\label{vbdef}
\end{equation}
where $\ell_b$ provides the length scale of the spatial variation.
Since $\sqrt{\mu-\bar{\mu}}>0$ is assumed finite and fixed,
$\ell_b\sim \ell$.  Of course, an analogous behavior occurs
on the opposite side, i.e. for $z = -z_b$.

The scaling behavior around the critical plane $z=z_b$ can be derived
using the same arguments of Ref.~\cite{DV-17}, applying them to the
particular case of slab geometries where the harmonic potential is
only applied along the transverse direction, while the system is
translationally invariant along the planar directions. The system
develops critical correlations around the planes $z=z_b$,
with a length scale
\begin{equation}
\xi_b\sim \ell_b^{\theta_b},\qquad \theta_b = {\nu\over 1 + \nu} = 0.40181(3).
\label{xielthetab}
\end{equation}
 For example, the one-particle correlation
function along a transverse direction is expected to scale as
\begin{equation}
G[({\bm x},z_1), ({\bm x},z_2)] \approx 
\xi_b^{-1-\eta} {\cal G}_b(\Delta z_1/\xi_b,
\Delta z_2/\xi_b).
\label{grscab}
\end{equation} 
Of course, such a scaling behavior at the critical planes is
anisotropic, distinguishing the planar and transverse
directions. However, both length scales along planar and transverse
directions are expected to scale as $\ell_b^{\theta_b}$.

\section{Low-temperature behavior of quasi-2D bosonic gases}
\label{lowtbh}

This section summarizes some exact results which are expected to
characterize the low-temperature QLRO phase of 
quasi-2D interacting bosonic
gases up to the BKT transition.

\subsection{The QLRO phase below the BKT transition}
\label{fssqlro}

The general universal features of the QLRO phase of quasi-2D systems
with a U(1) symmetry are described by the Gaussian spin-wave theory
\begin{equation}
H_{\rm sw} = {\beta\over 2} \int d^2 x \, (\nabla \varphi)^2.
\label{s2dxy}
\end{equation}
For $\beta\ge 2/\pi$, corresponding to $0\le \eta \le 1/4$, this
spin-wave theory describes the QLRO phase.  The values $\beta=2/\pi$
and $\eta=1/4$ correspond to the BKT transition~\cite{It-Dr-book}.

The spin-wave correlation function
\begin{equation}
G_{\rm sw}({\bm x}_1-{\bm x}_2) = 
\langle e^{-i\varphi({\bm x}_1)} e^{i\varphi({\bm x}_2)}\rangle
\label{gdefgau}
\end{equation}
is expected to provide the asymptotic large-$L$ behavior of the
two-point function of 2D interacting bosonic gases within the QLRO
phase.  For $|{\bm x}_1 - {\bm
  x}_2|\ll L$, 
\begin{equation}
G_{\rm sw}({\bm x}_1,{\bm x}_2)\sim {1\over |{\bm x}_1-{\bm x}_2|^{\eta}}, 
\label{gslx}
\end{equation}
where the exponent $\eta$ is related to the coupling $\beta$ by
\begin{equation}
\eta = {1\over 2\pi\beta}.
\label{etade}
\end{equation}

The general size dependence of $G_{\rm sw}$ on a square $L^2$ box
with PBC is also
known:~\cite{CNPV-13,CFT-book,It-Dr-book,Hasenbusch-05}
\begin{eqnarray}
&&G_{\rm sw}({\bm x},L) = C({\bm x},L)^\eta \times E({\bm x},L),\label{gsw} \\
&&C({\bm x},L) = {e^{\pi y_2^2} \theta'_1(0,e^{-\pi})\over |
\theta_1[\pi(y_1+iy_2),e^{-\pi}]|},
\nonumber \\
&&E({\bm x},L)=
{\sum_{n_1,n_2=-\infty}^{\infty} W(n_1,n_2) 
\cos[2\pi(n_1 x_1+n_2 x_2)]
\over \sum_{n_1,n_2=-\infty}^{\infty} W(n_1,n_2) },
\nonumber\\
&&W(n_1,n_2) = \exp[-\pi(n_1^2+n_2^2)/\eta],
\nonumber
\end{eqnarray}
where ${\bm x}\equiv (x_1,x_2)$, $y_i\equiv x_i/L$, $\theta_1(u,q)$
and $\theta_1'(u,q)$ are $\theta$ functions~\cite{Gradstein}.

Using Eq.~(\ref{gsw}), one can easily compute the universal function
$R_L(\eta)$, where $R_L\equiv \xi/L$ and $\xi$ is the
second-moment correlation length defined as
\begin{eqnarray}
&&\xi^2 =  {L^2\over 4 \pi^2} \left( {\chi \over \chi_1}-1\right),\\
&&\chi =  \int d^2x \,G_{\rm sw}({\bm x}),\quad
\chi_1 =  \int d^2x \,{\rm cos}\left({2\pi x_1\over L}\right) 
\,G_{\rm sw}({\bm x}).
\nonumber
\end{eqnarray}

Analogous results are obtained for the helicity
modulus~\cite{Hasenbusch-05}
\begin{eqnarray}
Y(\eta) = {1\over 2\pi\eta} - {\sum_{n=-\infty}^\infty n^2 \,{\rm
    exp}(-\pi n^2/\eta)\over \eta^2 \sum_{n=-\infty}^\infty {\rm exp}
  (-\pi n^2/\eta)}.
\label{rupstoro}
\end{eqnarray}

The above asymptotic large-$L$ behaviors (at fixed $T$ or $\eta$) are
approached with power-law corrections, indeed
\begin{eqnarray}
&& R_L(L,\eta)\equiv \xi/L = R_L(\eta) + a L^{-\varepsilon}, \label{fitXlt}\\
&& Y(L,\eta) = Y(\eta) + a L^{-\zeta}, \label{fitYlt}
\end{eqnarray}
respectively, where $\varepsilon$ and $\zeta$ are the exponents
associated with the expected leading corrections:\cite{PV-13,HPV-05}
\begin{eqnarray}
&&\varepsilon={\rm Min}[2-\eta,\kappa],\qquad
\zeta={\rm Min}[2,\kappa],\label{varezeta}\\
&&\kappa=1/\eta-4 + O[(1/\eta-4)^2].
\label{omega}
\end{eqnarray}

With increasing $T$ within the QLRO phase, the critical exponent
$\eta$ of the two-point function, cf. Eq.~(\ref{gslx}), increases up
to $\eta=1/4$ corresponding to the BKT transition.  Therefore, close
to the BKT transition, i.e. for $T\lesssim T_{\rm BKT}$, we may expand
the universal curves $R_L(\eta)$ and $Y(\eta)$ around $\eta=1/4$,
obtaining
\begin{eqnarray}
R_L(\eta) &=& 0.7506912222 + 1.699451\,\left( {1\over 4}-\eta\right) + \ldots,
\qquad
\label{rxitoroe}\\
Y(\eta) &=& 0.6365081782 +2.551196 \, \left( {1\over 4}-\eta\right) + \ldots
\label{ytoroe}
\end{eqnarray}

We expect that the above universal behaviors are also realized in the
low-temperature phase of BH models within slab geometries, for
$T<T_{\rm BKT}$, by the two-point functions $g({\bm x})$,
cf. Eq.~(\ref{gx1x2}), and the quantities $R_L\equiv \xi/L$ and $Y$
defined in Eq.~(\ref{Rdef}).

\subsection{Finite-size behavior at the BKT transition}
\label{fssBKT}

The BKT transition is characterized by logarithmic corrections to the
asymptotic behavior, due to the presence of marginal
renormalization-group (RG) perturbations at the BKT fixed
point~\cite{AGG-80,HMP-94,HP-97,Balog-01,Hasenbusch-05,PV-13}. 

The asymptotic behaviors at the BKT transition for $R_L$ and $Y$ can
be obtained by replacing~\cite{Hasenbusch-05,PV-13}
\begin{eqnarray}
1/4-\eta\approx {1\over 8 w},
\qquad w = \ln {L\over \Lambda} + {1\over2} \ln\ln {L\over \Lambda},
\label{replt}
\end{eqnarray}
into Eqs.~(\ref{rxitoroe}) and (\ref{ytoroe}).  The nonuniversal
details that characterize the model (such as the thickness $Z$ of the
quasi-2D BH models) are encoded in the model-dependent scale
$\Lambda$.  Thus one obtains the asymptotic large-$L$ behavior
\begin{eqnarray}
R(L,T_{\rm BKT}) = R^* + C_{R} w^{-1} + O(w^{-2}).
\label{rasym}
\end{eqnarray}
for both $R=Y,\,R_L$, with
\begin{eqnarray}
&Y^* = 0.6365081789,\quad & C_Y = 0.31889945,
\label{retalo}\\
&R_L^* = 0.7506912222, \quad & C_{R_L} = 0.21243137,
\label{rxilo}
\end{eqnarray}
for PBC.

In numerical analyses, Eq.~(\ref{rasym}) may be used to locate the BKT
transition point, i.e. by requiring that the finite-size dependence of
the data matches it.  However we note that this straightforward
approach is subject to systematic errors which get suppressed only
logarithmically with increasing $L$.  This makes the accuracy of the numerical or
experimental determination of the critical parameters quite problematic.
This problem can be overcome by the so-called matching
method~\cite{HMP-94,HP-97,Hasenbusch-05,Hasenbusch-08,Hasenbusch-09,
  Hasenbusch-12,CNPV-13}, which allows us to 
control the whole pattern of the logarithmic corrections, leaving only
power-law corrections. 

The matching method exploits the fact that the finite-size behavior of
RG invariant quantities $R$, such as $R_L$ and $Y$, of different
models at their BKT transition shares the same logarithmic corrections
apart from a nonuniversal normalization of the scale.  Indeed, the
$L$-dependence of two models at their BKT transition
is related by the asymptotic relation
\begin{equation}
R^{(1)}(L_1, T_{\rm BKT}^{(1)}) \approx 
R^{(2)}(L_2=\lambda L_1, T_{\rm BKT}^{(2)}) ,
\label{r12rel}
\end{equation}
apart from power-law corrections, which are $O(L^{-2})$ for the
helicity modulus $Y$ and $O(L^{-7/4})$ for the ratio $R_L$.  
The matching parameter $\lambda$
is the only free parameter, but it does not depend on the
particular choice of the RG invariant quantity.  The matching method consists in
finding the optimal value of $T$ matching the finite-size behavior of
$Y$ and $R_L$ of the 2D $XY$ model whose value of $T_{\rm BKT}$ is
known with high accuracy~\cite{HP-97,Hasenbusch-05}.  The complete
expression of $R_L$ and $Y$ of the 2D $XY$ model have been numerically
obtained by high-precision numerical
studies~\cite{Hasenbusch-05,Hasenbusch-08} and by extrapolations using
RG results for the asymptotic behavior.  For example, 
the $L$-dependence of the helicity modulus $Y$ at the BKT transition of
the 2D $XY$ model is accurately
reconstructed by the following expression~\cite{CNPV-15}
\begin{eqnarray}
\widetilde{Y}_{XY}(L) & \equiv & Y_{XY}(T_{\rm BKT},L) = 
\label{eqytilde} \\         
&=& 0.6365081782 
+ 0.318899454 \,w^{-1}  \nonumber\\
&+& 2.0319176\, w^{-2}  - 40.492461 \, w^{-3}  
\nonumber\\
&+& 325.66533 \,w^{-4}
- 874.77113 \,w^{-5} 
\nonumber\\
&+& 8.43794\,L^{-2} + 79.1227\,L^{-4} 
- 210.217\,L^{-6},  \nonumber
\end{eqnarray}
where $w$ is given in Eq.~(\ref{replt}) with
$\Lambda=\Lambda_{XY}=0.31$.  

The matching method has been already applied~\cite{CNPV-13} to the 2D BH
models (\ref{bhm}), obtaining the accurate estimate $T_{\rm
  BKT}=0.6877(2)$ in the hard-core $U\to\infty$ limit and at half
filling ($\mu=0$).

\section{Off-equilibrium slow dynamics  and  dimensional crossover}
\label{KZ}

The dynamical behavior of statistical systems driven across phase
transitions is a typical off-equilibrium phenomenon.  Indeed, the
large-scale modes present at the transition are unable to reach
equilibrium as the system changes phase, even when the time scale
$t_s$ of the variation of the system parameters is very large.  Such
phenomena are of great interest in many different physical contexts,
at both first-order and continuous transitions, where one may observe
hysteresis and coarsening phenomena, the Kibble-Zurek (KZ) defect
production, etc, see,
e.g., Refs.~\cite{Kibble-76,Zurek-85,Binder-87,Bray-94,CG-05,GZHF-10,
PSSV-11,CEGS-12,
  NGSH-15,Braun-etal-15,Biroli-15,PV-16,DWGGP-16,ARBBHC-16,BN-17,PV-17}.
 The correlation functions  obey general off-equilibrium scaling (OS)
laws in the limit of  large time scale $t_s$ of the variations
across the transition, which are controlled by the universal static and dynamic
exponents of the equilibrium transition~\cite{GZHF-10,CEGS-12,PV-17}.

We now consider the off-equilibrium behavior arising from slow time
variations of the temperature $T$ across the BEC transition.  
We assume a standard linear protocol, varying $T$ so that
\begin{equation}
\delta(t)\equiv 1 - T(t)/T_c = t/t_s,
\label{deltat}
\end{equation} 
starting at a time $t_i<0$ in the high-$T$ phase and ending at $t_f >
0$ in the low-$T$ phase. $t_s$ is the time scale of the temperature
variation. The BEC transition point corresponds to $t=0$ (however this
is not strictly required, it is only convenient for our discussion).
Several experiments implementing off-equilibrium time-dependent
protocols in cold-atom systems have been reported, see, e.g.,
Refs.~\cite{SHLVS-06,WNSBD-08,LDSDF-13,NGSH-15,CCBDWNBD-15,BN-17}.

Beside the static critical exponent~\cite{CHPV-06} $\nu=0.6717(1)$ of
the 3D $XY$ universality class, we also need information on the critical
dynamic behavior at the BEC transition. This is characterized by the
dynamic exponent $z=d/2$, thus $z=3/2$ in 3D, associated with the
model-F dynamics~\cite{HH-77,FM-06} which is conjectured to describe
the dynamic universality class of the 3D BEC transition.

In the standard thermodynamic limit of cubic-like boxes, with $L_1\sim
L_2 \sim L_3 \sim L$ and $L\to\infty$, one defines the OS limit as the
large-time-scale limit, $t_s\to\infty$, keeping the OS scaling variables
\begin{equation}
{\cal T} \equiv t/t_s^{\kappa},\qquad {\bm x}_s \equiv {\bm x}/t_s^\zeta,
\label{txkz}
\end{equation}
fixed. Scaling arguments allow us to determine the appropriate
exponents $\kappa$ and $\zeta$, obtaining~\cite{Kibble-76,Zurek-85,CEGS-12}
\begin{equation}
\kappa = {z\nu\over 1 + z \nu},\qquad \zeta = {\nu\over 1 + z\nu},
\label{kappaexp}
\end{equation}
where $\nu$ and $z$ are the static correlation-length and dynamic
exponents. In particular, by inserting the values of $\nu$ and $z$, we
obtain $\kappa = 0.50188(4)$ and $\zeta = 0.33459(3)$.  

We may apply these OS arguments to the equal-time two-point
correlation function, measured after a time $t$ and averaged over the
initial Gibbs distribution at a given initial temperature $T>T_c$.  Standard
scaling arguments lead to the OS asymptotic behaviors~\cite{CEGS-12}
\begin{equation}
G({\bm x},t,t_s) \approx t_s^{-\zeta(1+\eta)} {\cal G}_o({\bm x}_s,{\cal T}).
\label{gxkxs}
\end{equation}
Moreover, we expect
\begin{equation}
\xi(t,t_s) \approx t_s^{\zeta} \,{\cal R}_o({\cal T}),
\label{kzscalingxi}
\end{equation}
for any length scale associated with the critical modes.  Experimental
studies of this dynamic behavior, and the related KZ defect
production, led to the estimate~\cite{NGSH-15} $\zeta=0.35(4)$, which
is in good agreement with the theoretical result (\ref{kappaexp}).

We now discuss how this off-equilibrium behavior may change in quantum
gases confined within slab geometries with $Z\ll L$, and in particular
with a finite thickness $Z$ and infinite $L\to\infty$ planar sizes.
Analogous experiments with quasi-2D cold-atom systems constrained in
slab geometries have been reported in Refs.~\cite{CCBDWNBD-15,BN-17}
(homogenous hard-wall traps along the planar directions and harmonic
along the transverse direction).  They observe the emergence of
coherence when cooling the atomic gas through the BEC temperature.

The off-equilibrium behavior arising from the slow variation of the
temperature across the BEC transition point is made particularly
complex by the presence of a close quasi-2D BKT transition. Thus,
disentangling the behaviors corresponding to BEC and BKT is quite hard
in experimental or numerical analyses.  The authors of
\cite{CCBDWNBD-15,BN-17} interpreted the observed behavior as a
transverse condensation
phenomenon~\cite{CCBDWNBD-15,BN-17,vDK-97,AJKB-11}.  In the following
we put forward an alternative framework to describe the dimensional
crossover in slab geometries, based on an off-equilibrium FSS (OFSS).

As already said, for a finite thickness $Z$, even though $L\to\infty$,
the system does not develop a diverging correlation length at the 3D
BEC transition temperature $T_c$, but $\xi$ remains of the order of
the transverse size $Z$. Thus the systems can evolve adiabatically,
i.e. its evolution can be perfomed by passing through
quasi-equilibrium states for a sufficiently large time scale $t_s$ of
the variation of $T(t)$ around $T_c$.  This is possible until it
reaches the BKT transition at the time $t>0$ corresponding to $T_{\rm
  BKT}$, i.e. when $t/t_s= \delta_{\rm BKT} \equiv 1 - T_{\rm
  BKT}/T_c$. 

Of course, the OS at the BKT transition is expected to substantially
differ from that at the 3D BEC transition, such as Eqs.~(\ref{gxkxs})
and (\ref{kzscalingxi}), because it must be controlled by the 2D
universality class of the BKT transition in quantum gases.  At the BKT
transition the relevant exponents for KZ off-equilibrium protocols are
expected to be $\nu=\infty$ (related to the exponential increase of
the correlation length when $T\to T_{\rm BKT}^+$) and $z=1$ (2D model
F of the dynamics). Thus the power laws of the off-equilibrium scaling
variables (\ref{txkz}) at the BKT transition lead to $\kappa=1$, apart
from logarithms.

However things become quite involved when the thickness $Z$ becomes
large because the BKT transition gets very close to the BEC
temperature $T_c$, cf. Eq.~(\ref{deltabkt}). Therefore, the analysis
of numerical and experimental data may become hard, and
straightforward power-law fits may turn out to be misleading. 

In order describe the time-dependent dimensional crossover of slab
geometries under the protocol (\ref{deltat}), we consider an OFSS
framework involving the size $Z$ of the transverse direction.  The
appropriate OFSS limit is defined by introducing the scaling variables
\begin{equation}
X_o = Z^{1/\nu} \delta(t),\qquad W_o= Z^{-1/\zeta} t_s,
\label{abscal}
\end{equation}
In the OFSS limit a length scale, such as the planar correlation
length defined in Eq.~(\ref{eq:xi_definition}), is expected to behave
as
\begin{equation}
\xi(Z,t,t_s)  \approx Z \; {\cal S}_o(X_o,W_o),
\label{scalxiab}
\end{equation} 
where ${\cal S}_o$ is a universal OFSS function.  

In this OS framework the equilibrium FSS around $T_c$ is recovered in
the limit $W_o\to\infty$, i.e.
\begin{equation}
{\cal S}_o(X_o,W_o\to\infty) = {\cal R}(X_o)
\label{soro}
\end{equation}
where ${\cal R}(X_o)$ is the equilibrium FSS function,
cf. Eq.~(\ref{crosslimit}).  In particular, at $t=0$ corresponding to
$T(t)=T_c$, we expect to recover the equilibrium result $\xi\sim Z$
when $t_s\gg Z^{1/\zeta}$.  Note however that the equilibrium limit
is not well defined for any $X_o$, because it diverges when $X_o
\ge X_{\rm BKT}$, cf. Eq.~(\ref{xbkt}), corresponding to the BKT
transition. Around $X_{\rm BKT}$ the behavior of the scaling functions
must somehow show the off-equilibrium singularities associated with a
slow passage thorough a BKT transition.

The above scaling behaviors can be straightforwardly extended to the
case of a transverse harmonic trap, using the same TTSS arguments of
Sec.~\ref{tssslab}.  Apart from replacing $Z$ with $\ell^\theta$, the
main features of the OS behavior remain the same.

We mention that experiments under analogous time-dependent protocols
crossing the BEC transition 
have been performed with atomic gases confined in slab-like traps with
a transverse harmonic trapping
potential~\cite{CCBDWNBD-15,BN-17}. They were able to check the
initial 3D behavior, without a clear identification of the subsequent
quasi-2D behavior. The computation of the defect production arising
from the Kibble-Zurek mechanism is further complicated by later-time
coarsening phenomena~\cite{Bray-94,CEGS-12,Biroli-15}.

\section{Numerical results for the BH model}
\label{dimcross}

\begin{figure}
\includegraphics[width=7.5cm]{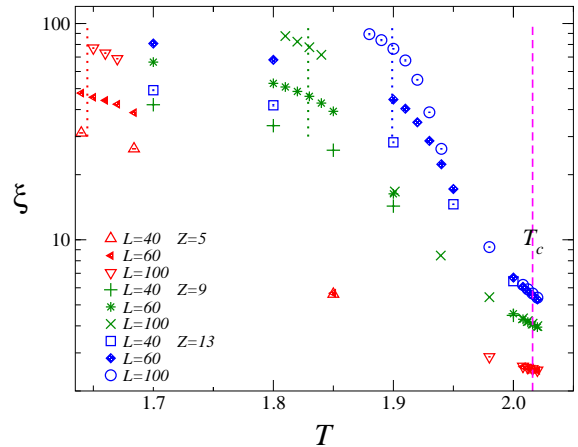}
\caption{ QMC data of the planar correlation length $\xi$ for the
  hard-core $U\to\infty$ BH model at zero chemical potential, for
  various values of $Z$, $L$ and $T$.  The dashed vertical line
  indicates the BEC transition temperature $T_c$, the dotted vertical
  lines indicate our estimates of the BKT transition temperature for
  $Z=5,\,9,\,13$ (from the left to right).  The statistical errors of
  the data are so small to be hardly visible.  }
\label{xidata}
\end{figure}

In order to check the dimensional crossover scenario discussed in the
previous section, we present a numerical study of the equilibrium
properties of the BH model (\ref{bhm}) in the hard-core $U\to\infty$
limit and at zero chemical potential $\mu=0$, corresponding to half
filling, i.e., $\langle n_{\bm r} \rangle = 1/2$ for any $T$. In the
hard-core limit and for $\mu=0$, the 3D BEC transition occurs at
$T_c=2.01599(5)$ and the 2D BKT transition at $T_{\rm BKT}=0.6877(2)$.

Numerical results are obtained by quantum Monte Carlo (QMC)
simulations using the directed operator-loop algorithm
\cite{SK-91,SS-02,DT-01}.  We consider slab geometries, i.e.
$L^2\times Z$ lattices with $Z\ll L$, with OBC along the transverse
directions, and PBC along the planar directions. We present numerical
results for some values of the thickness $Z$, in particular
$Z=5,\,9,\,13$, various planar sizes up to $L\approx 100$, and several
values of the temperature $T\lesssim T_c$.  The maximum size $Z$ of
our numerical study is limited by the fact that the computational
effort of QMC rapidly increases, because they also require larger
values of the planar sizes.

\begin{figure}
\includegraphics[width=7cm]{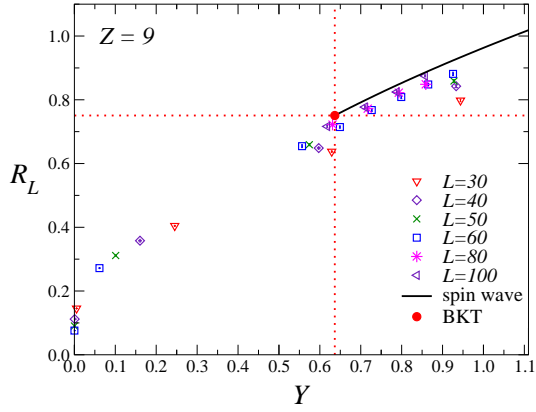}
\vskip10mm
\includegraphics[width=7cm]{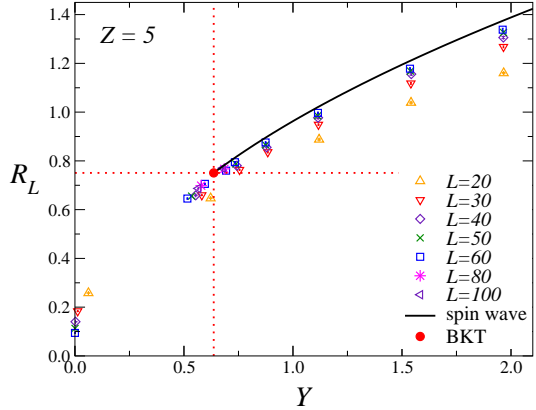}
\caption{ $R_L\equiv \xi/L$ versus $Y$ for $Z=5$ (bottom) and $Z=9$
  (top), and for several values of $L$ and $T$.  The full line shows
  the spin-wave curve $R_L(Y)$ which is expected to be asymptotically
  approached for $L\to\infty$ within the QLRO phase; its end point
  corresponds to the BKT transition.  In particular, for $Z=5$ the
  values of $T$ of the data shown in the bottom figure are (from right
  to left) $T=$ 1.1858, 1.3518, 1.5179, 1.6, 1.64, 1.65, 1.67, 1.6839,
  1.85.  The behavior of the data close to the BKT point suggests
  $T_{\rm BKT}(Z=5)\approx 1.65$.  Analogously for $Z=9$ the data are
  for $T=$1.8, 1.81, 1.82, 1.83, 1.84,1.85, 1.9, 2; they suggest that
  $T_{\rm BKT}(Z=9)\approx 1.83$.  The statistical errors of the data
  are so small to be hardly visible.  }
\label{yurxi}
\end{figure}

\begin{figure}
\includegraphics[width=7cm]{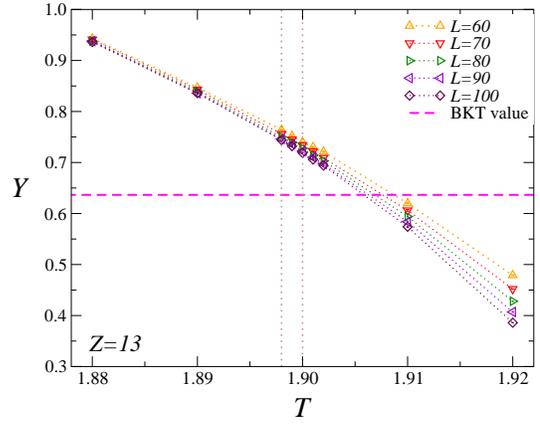}
\vskip10mm
\includegraphics[width=7cm]{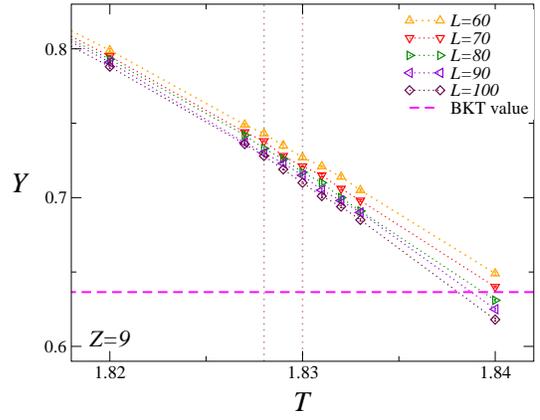}
\caption{ Data of $Y$ for $Z=9$ (bottom) and $Z=13$ (top) around the
  corresponding BKT temperatures.  Analogous results have been
  obtained for $Z=5$. The dashed horizontal line indicates the BKT
  value $Y^*=0.6365...$. The dotted vertical lines indicate the
  interval corresponding to our best estimates of $T_{\rm BKT}$,
  i.e. $T_{\rm BKT}=1.829(1)$ for $Z=9$ and $T_{\rm BKT}=1.899(1)$ for
  $Z=13$, obtained by the matching procedure.  }
\label{yute}
\end{figure}

We compute the observables defined in Sec.~\ref{moobs}.  In QMC
simulations the helicity modulus is obtained from the linear winding
number $W_a$ along the $a^{\rm th}$ direction, i.e.
\begin{eqnarray}    
Y\equiv Y_a = \langle W_a^2 \rangle,\qquad
W_a = {N_a^+ - N_a^-\over L}
\label{winding}
\end{eqnarray}
where $N_a^+$ and $N_a^-$ are the numbers of non-diagonal operators
which move the particles respectively in the positive and negative
$a^{\rm th}$ direction.

Figure \ref{xidata} shows data for the planar second-moment
correlation length $\xi$ defined in Eq.~(\ref{eq:xi_definition}), for
$Z=5,\,9,\,13$ and $T\lesssim T_c$.  We observe that $\xi$ is small
for $T>T_c$, and apparently $L$- and $Z$-independent (for sufficiently
large $L$ and $Z$), indicating that it remains finite in the
large-$L$ and large-$Z$ limit.  Around $T_c$ the data of $\xi$ appear
to converge to a finite value when increasing $L$ at fixed $Z$;
however, they show that $\xi$ increases with increasing $Z$,
approximately as $\xi\sim Z$.  Then, for sufficiently small values of
$T$, the data begin showing a significant dependence on $L$. At low
temperature we observe $\xi\sim L$ at fixed $T$, suggesting that $\xi$
diverges with increasing $L$ even when keeping $Z$ fixed.  In the
following we show that this apparently complicated behavior can be
explained by the dimensional crossover scenario put forward in the
previous sections.

To begin with, we investigate the nature of the low-temperature
behavior where the planar correlation length $\xi$ appears to diverge with increasing $L$.
According to the arguments of the previous sections, at low
temperature BH systems for any thickness $Z$ should show a quasi-2D
QLRO phase, whose behavior is essentially described by the 2D
spin-wave theory, see in particular Sec.~\ref{lowtbh}.  As discussed
in Sec.~\ref{fssqlro}, this implies universal relations among the
ratio $R_L\equiv \xi/L$, the quasi-2D helicity modulus $Y$ and the
exponent $\eta$ characterizing the planar two-point correlation
function. In Fig.~\ref{yurxi} we plot data of $R_L$ versus those of
$Y$, comparing them with the universal curve $R_L(Y)$ which can be
easily obtained from the spin-wave results reported in
Sec.~\ref{fssqlro}. This curve ends at the BKT point
$(Y^*,R_L^*)=(0.6365..., 0.7506...)$.  For sufficiently small $T$,
depending on the value of $Z$, the data approach the universal
spin-wave curve $R_L(Y)$ with increasing $L$.  Extrapolations using
the expected power-law corrections, cf. Eqs.~(\ref{fitXlt}) and
(\ref{fitYlt}), turn out to be consistent with the exact spin-wave
results. Therefore, the numerical results nicely support the existence
of a QLRO phase for any $Z$, with the expected universal spin-wave
behaviors.

\begin{figure}
\includegraphics[width=7cm]{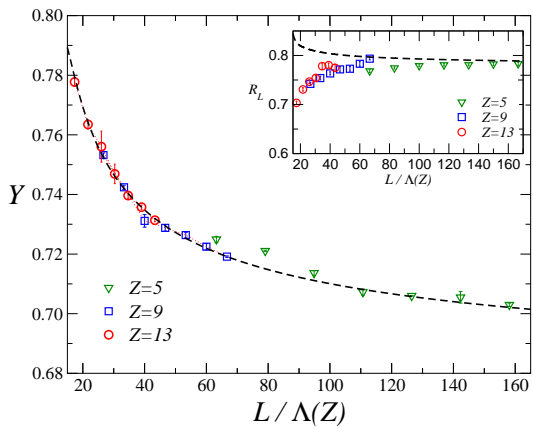}
\caption{ Finite-size dependence of $Y$ at the BKT transition. We plot
  data for $Z=5,\,9,\,13$ at their best estimates of $T_{\rm BKT}$,
  i.e. $T=1.645,\,1.829,\,1.899$ respectively, versus $L/\Lambda(Z)$
  with $\Lambda(Z)=\Lambda_{XY}/\lambda(Z)=0.6,\,1.5,\,2.4$
  respectively. The dashed line shows the $XY$ curve (\ref{eqytilde}).
  The inset shows data of $R_L$, with the corresponding
  $XY$ curve (dashed line) taken from Ref.~\cite{CNPV-13}. 
}
\label{tbktscal}
\end{figure}

We also note that above a given temperature, depending on the
thickness $Z$, the data do not  approach the spin-wave
curve $R_L(Y)$ anymore, as it is expected to occur for $T>T_{\rm BKT}$
where both $R_L$ and $Y$ vanish in the large-$L$ limit.  Therefore,
the data of Fig.~\ref{yurxi} allow us to approximately locate $T_{\rm
  BKT}$ between the temperature values of the data closest to the BKT
point $(Y^*,R_L^*)$ which respectively approach the spin-wave curve
and deviate from it.  We already note that $T_{\rm BKT}$ increases
with increasing $Z$.  This can be also inferred by the data of the
helicity modulus $Y$ versus the temperature, see Fig.~\ref{yute}.
They are generally decreasing, and for sufficiently large $T$ they
appear to cross the value $Y=Y^*\approx 0.6365$ corresponding to the
BKT transition, indicating that those values of $T$ are larger than
$T_{\rm BKT}$.

More accurate estimates of $T_{\rm BKT}$ can be obtained by looking
for the optimal values of $T$ achieving the matching of the available
data of $Y$ and $R_L$ with the finite-size dependence 
of the 2d $XY$ model at its BKT transition,
see Sec.~\ref{fssBKT}. In particular, $T_{\rm BKT}(Z)$
is given by the value of $T$ providing the optimal matching 
of the data of $Y(Z,L,T)$ with the
finite-size dependence of the helicity modulus of the 
2D $XY$ model, i.e.
\begin{equation}
Y(Z,L,T) = \widetilde{Y}_{XY}[\lambda(Z) L] + O(L^{-2}),
\label{yzmatching}
\end{equation}
with $\widetilde{Y}_{XY}$ given by Eq.~(\ref{eqytilde}).  Some
matching procedures are described in Ref.~\cite{CNPV-13}.  This
numerical analysis largely suppresses the systematic error, because it
is not affected by logarithmic corrections, but only $O(L^{-2})$
power-law corrections.  For $Z=1$ the optimal matching led to the
estimate $T_{\rm BKT}(Z=1)=0.6877(2)$ and $\lambda(Z=1) \approx 1.5$.

We determine the optimal values of $T$ and $\lambda(Z)$ satisfying the
scaling relation (\ref{yzmatching}).  We skip most details of the
numerical matching procedures, which can be found in
Ref.~\cite{CNPV-13}. We only mention that we use QMC data from $L=20$
to $L=100$, for sufficiently close values of $T$ to obtain reliable
estimates for any $T$ by interpolation, see Fig.~\ref{yute}.  Our
estimates for the optimal matching parameters are $T_{\rm
  BKT}(Z=5)=1.645(2)$, $T_{\rm BKT}(Z=9)=1.829(1)$, and $T_{\rm
  BKT}(13)=1.899(1)$; correspondingly we obtain $\lambda(Z=5)=0.4(2)$,
$\lambda(Z=9)=0.20(5)$, $\lambda(Z=13)=0.14(2)$.  The statistical
error of the analysis is estimated using bootstrap methods. The error
reported above takes also into account the variations of the results
when changing the procedure to obtain the optimal matching, for
example when considering or not the $O(L^{-2})$ scaling corrections,
and varying the minimum size $L$ of the data used in the
analysis. 

The quality of the matching can be inferred from Fig.~\ref{tbktscal},
which shows the data at the optimal matching values of $T_{\rm BKT}$
versus the ratio $L/\Lambda(Z)$ with $\Lambda(Z) =
\Lambda_{XY}/\lambda(Z)$, so that all data of $Y$, for any $Z$, are
expected to follow the same curve $\widetilde{Y}_{XY}$ versus
$L/\Lambda_{XY}$ with $\Lambda_{XY}=0.31$.  This is indeed what we
observe, apart from some scaling corrections at the smallest values of
$L$, which are expected to get suppressed as $O(L^{-2})$.  We consider
the results of the matching analysis of the $Y$ data as our best estimates of $T_{\rm
  BKT}$.  Note also that the values of $\lambda(Z)$ are decreasing, as
expected because the value $\lambda(Z) L$ is somehow related to the
equivalent planar size of the lattice, and for slab geometries one may
expect that this is approximately given by the aspect ratio $L/Z$,
thus $\lambda(Z)\sim 1/Z$ roughly.

An analogous numerical analysis can be done using the data of $R_L$.
However it turns out to be less accurate due to larger scaling
corrections.  As also observed in Ref.~\cite{CNPV-13}, $R_L$ is
subject to significantly large power-law scaling corrections, which
decrease as $L^{-7/4}$.  The $XY$ curve of $R_L$ is reported in
Ref.~\cite{CNPV-13}.  Note that once determined $T_{\rm BKT}$ and
$\lambda(Z)$, there are no other free parameters to optimize the
matching.  The inset of Fig.~\ref{tbktscal} shows the data and their
comparison with the $XY$ curve using the values of $T_{\rm BKT}$ and
$\lambda(Z)$ obtained from the analysis of the data of $Y$. The data
appear to approach the asymptotic curve with increasing $L$, therefore
they are consistent with the theoretical predictions.  However, as
already mentioned, we note that the approach to the expected
asymptotic behavior is characterized by larger scaling corrections,
thus requiring larger lattice sizes to obtain independent estimates of
$T_{\rm BKT}$ as accurate as those obtained using the data of $Y$.

\begin{figure}
\includegraphics[width=7cm]{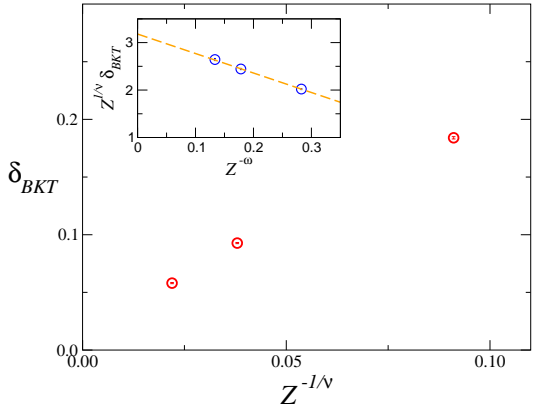}
\caption{ Estimates of $\delta_{\rm BKT}(Z)\equiv 1 - T_{\rm
    BKT}(Z)/T_c$ vs $Z^{-1/\nu}$ with $\nu=0.6717$.  The data are
  compatible with the expected behavior $\delta_{\rm BKT}(Z)\sim
  Z^{-1/\nu}$.  The inset shows the product $Z^{1/\nu}\delta_{\rm
    BKT}(Z)$ versus $Z^{-\omega}$ with $\omega=0.785$ (the dashed line
  is obtained by a linear fit), which is the expected behavior of the
  leading scaling corrections.  }
\label{devsZ}
\end{figure}

Figure \ref{devsZ} shows $\delta_{\rm BKT}(Z) \equiv 1 - T_{\rm
  BKT}(Z)/T_c$ versus $Z^{-1/\nu}$, as obtained from the above
estimates of $T_{\rm BKT}$. The data turn out to be consistent with
the expected asymptotic behavior $\delta_{\rm BKT}(Z)\sim Z^{-1/\nu}$.
We also estimate
\begin{equation}
X_{\rm BKT} = {\rm lim}_{Z\to\infty} Z^{1/\nu} \delta_{\rm BKT} =
3.2(1),
\label{xbktest}
\end{equation}
by extrapolating the available data for the product
$Z^{1/\nu}\delta_{\rm BKT}$ using the ansatz
\begin{equation}
Z^{1/\nu} \delta_{\rm BKT} = X_{\rm BKT} + c \,Z^{-\omega},
\label{ansatz}
\end{equation}
see the inset of Fig.~\ref{devsZ}, where $\omega=0.785(20)$ is the
leading scaling-correction exponent of the 3D $XY$ universality class.

\begin{figure}
\includegraphics[width=7cm]{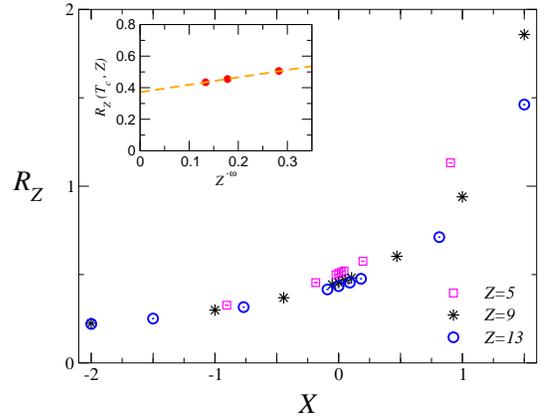}
\caption{ Scaling of $R_Z$ around $T_c$ (most data are taken for
  $L=120$, which is sufficiently large to provide a good approximation
  of the $L\to\infty$ limit in the range of $Z$ and $X$ values
  considered, within about 1\%.  We plot the data versus $X\equiv
  Z^{1/\nu} \delta$ with $\nu=0.6717$ and $\delta\equiv 1- T/T_c$.
  The inset shows the data of $R_Z$ at $T_c$ versus $Z^{-\omega}$
  which is the expected behavior of the leading scaling corrections
  (the dashed line is obtained by a linear fit).  }
\label{checkscatc}
\end{figure}

\begin{figure}
\includegraphics[width=7cm]{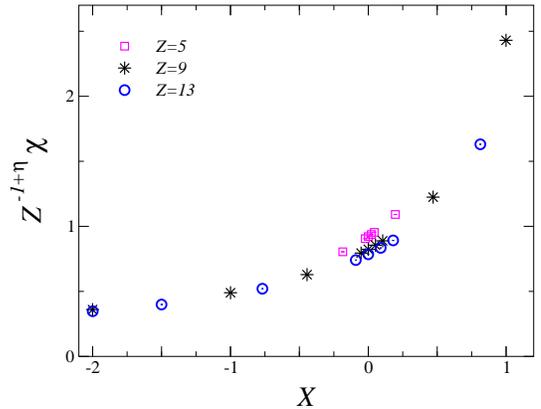}
\caption{ Scaling of the planar susceptibility $\chi$ around $T_c$.
  We plot $Z^{-1+\eta} \chi$ versus $X\equiv Z^{1/\nu} \delta$. 
The data support the asymptotic TFSS $\chi\approx Z^{1-\eta} f_\chi(X)$. }
\label{checkscatcchi}
\end{figure}

Finally, we check the TFSS $R_Z\approx f_\xi(X)$ with
$X=Z^{1/\nu}\delta$ around $T_c$, in the planar {\rm thermodynamic}
limit, i.e. when $\xi, Z\ll L$.  As argued in Sec.~\ref{fssslab}, the
scaling function $f_\xi(X)$ is expected to have an essential
singularity at $X_{\rm BKT}\approx 3.2$, cf. Eq.~(\ref{fxixs}).  In
Fig.~\ref{checkscatc} we show data of $R_Z$ around $T_c$ versus
$X\equiv \delta Z^{1/\nu}$.  They support the TFSS behavior of
$R_Z$. Scaling corrections are expected to decrease as $Z^{-\omega}$.
They appear significantly larger for $X>0$, when approaching the
singularity at $X_{\rm BKT}$.  By extrapolating the available data at
$T_c$ using $R_Z(Z,T_c) = R_Z^* + c \,Z^{-\omega}$ (see the inset of
Fig.~\ref{checkscatc}), we estimate $R_Z^*= 0.372(3)$ for the
universal large-$Z$ ratio $R_Z\equiv\xi/Z$ characterizing the TFSS of
the critical planar correlation length.  An analogous scaling behavior
is expected for the planar susceptibility defined as in
Eq.~(\ref{chi}).  The data shown in Fig.~\ref{checkscatcchi} nicely
support the corresponding TFSS (\ref{chitfss}).

\section{Summary}
\label{conclu}

We have studied the phase-coherence properties of Bose gases confined
within slab-like boxes of size $L^2\times Z$ with $Z\ll L$, at the 3D
BEC transition temperature $T_c$ and at lower temperatures.  Unlike
systems confined within cubic-like geometries, i.e. boxes with $L\sim
Z$, the low-temperature behavior of gases confined within slab
geometries is also characterized by the possibility of undergoing a
finite-temperature quasi-2D BKT transition at $T_{\rm BKT}<T_c$ with
$T_{\rm BKT}$ depending on the thickness $Z$. Below $T_{\rm BKT}$ the
planar one-particle correlations decay algebraically, as predicted by
the QLRO of the 2D spin-wave theory.

Therefore, Bose gases in slab geometries experience a dimensional
crossover with decreasing $T$, from 3D behaviors for $T\gtrsim T_c$ to
a quasi-2D critical behavior for $T\lesssim T_{\rm BKT}$.  However, in
the limit of large thickness $Z$ the quasi-2D BKT transition
temperature approaches that of the 3D BEC transition, i.e.  $T_{\rm
  BKT}\to T_c$ for $Z\to \infty$.  The interplay of 3D and quasi-2D
critical modes can be described by the TFSS limit for systems on slab
geometries: $Z\to\infty$ and $T\to T_c$ keeping the product
$(T-T_c)Z^{1/\nu}$ fixed (the planar sizes are assumed to be
infinite), where $\nu$ is the correlation-length exponent at the 3D
BEC transition.  The corresponding TFSS functions present an essential
singularity due to the quasi-2D BKT transition below $T_c$. A similar
TTSS behavior is also put forward in the case the particles are
trapped by a transverse harmonic potential in the limit of large
transverse trap size $\ell$. In the TTSS framework the length scale
$\xi_t=\ell^\theta$, where $\theta=2\nu/(1+2\nu)=0.57327(4)$, plays
the same role of the transverse size $Z$ of the TFSS.

We also extend the discussion to the off-equilibrium behavior arising
from slow time variations of the temperature $T$ across the BEC
transition. In particular we consider the linear protocol
$\delta(t)\equiv 1 - T(t)/T_c = t/t_s$ where $t_s$ is a time scale.
The corresponding off-equilibrium behavior is made particularly
complex by the presence of the close quasi-2D BKT transition at
$T_{\rm BKT}<T_c$, which is also crossed during the time-dependent
protocol.  Thus, disentangling the behaviors corresponding to BEC and
BKT is quite hard in experimental or numerical analyses. We argue that
the off-equilibrium behavior in the limit of large $t_s$ can be
described by an off-equilibrium FSS theory for bosonic gases confined
within slab geometries, extending the TFSS of the equilibrium
properties.

To provide evidence of the dimensional-crossover scenario in
interacting bosonic gases, we present a numerical study of the BH
model (\ref{bhm}) in anisotropic slab-like lattices $L^2\times Z$ with
$Z\ll L$. With decreasing $T$ from the high-temperature normal phase,
we first observe a quasi-BEC transition where the critical length
scale $\xi$ gets large, but it does not diverge, being limited by
$\xi\sim Z$ (keeping $Z$ fixed).  Then a BKT transition occurs to a
QLRO phase, where the system develops planar critical correlations
essentially described by the 2D Gaussian spin-wave theory.  We show
that the 3D$\to$2D dimensional-crossover scenario explains the
apparently complex dependence on $T$, $Z$, and $L$ of the one-particle
correlation functions and the corresponding length scale, when
decreasing the temperature from $T>T_c$ to $T < T_{\rm BKT}<T_c$.  The
results turn out to be consistent with the predictions of the TFSS at
the BEC transition.

The dimensional-crossover scenario is expected to apply to any quantum
gas of interacting bosonic particles constrained in boxes or lattice
structures with slab geometries.  Analogous arguments apply to $^4$He
systems in film geometries~\cite{GKMD-08}, and to 3D $XY$ spin models
defined in lattices with slab
geometries~\cite{SM-95,SM-96,SM-97,Hasenbusch-09}.

We conclude stressing that the above issues related to the
dimensional-crossover scenario are of experimental relevance since
cold-atom systems confined within slab geometries can be effectively
realized, see e.g.
Refs.~\cite{CCBDWNDB-14,NGSH-15,CCBDWNBD-15,BN-17}.  These
experimental setups offer the possibility of investigating the
dependence of the phase-coherence properties on the geometry of the
cold-atom system. Our study provides a framework to interpret the
experimental or numerical data related to the 3D$\to$2D dimensional
crossover in Bose gases confined within slab geometries, and in
particular their complicated dependence on the thickness $Z$.


\begin{thebibliography}{99}


\bibitem{CWK-02} E. A. Cornell and C. E. Wieman, Nobel Lecture:
  Bose-Einstein condensation in a dilute gas, the first 70 years and
  some recent experiments, Rev.\ Mod.\ Phys.\ {\bf 74}, 875 (2002);
  N. Ketterle, Nobel lecture: When atoms behave as waves:
  Bose-Einstein condensation and the atom laser,
  Rev.\ Mod.\ Phys.\ {\bf 74}, 1131 (2002).


\bibitem{Andrews-etal-97} M. R. Andrews C. G. Townsend, H.-J. Miesner,
  D. S. Durfee, D. M. Kurn, and W. Ketterle, Observation of
  Interference Between Two Bose Condensates, Science {\bf 275}, 637
  (1997).

\bibitem{Stenger-etal-99} J. Stenger, S. Inouye, A. P. Chikkatur,
  D. M. Stamper-Kurn, D. E. Pritchard, and W. Ketterle, Bragg
  Spectroscopy of a Bose-Einstein Condensate, Phys. Rev. Lett. {\bf
    82}, 4569 (1999).

\bibitem{Hagley-etal-99} E. W. Hagley, L. Deng, M. Kozuma,
  M. Trippenbach, Y. B. Band, M. Edwards, M. Doery, P. S. Julienne,
  K. Helmerson, S. L. Rolston, and W. D. Phillips, Measurement of the
  coherence of a Bose-Einstein condensate, Phys. Rev. Lett. {\bf 83},
  3112 (1999).

\bibitem{BHE-00} I. Bloch, T.W. H\"ansch, and T. Esslinger,
  Measurement of the spatial coherence of a trapped Bose gas at the
  phase transition, Nature {\bf 403}, 166 (2000).


\bibitem{Dettmer-etal-01} S. Dettmer, D. Hellweg, P. Ryytty,
  J. J. Arlt, W. Ertmer, K. Sengstock, D. S. Petrov,
  G. V. Shlyapnikov, H. Kreutzmann, L. Santos, and M. Lewenstein,
  Observation of Phase Fluctuations in elongated Bose-Einstein
  Condensates, Phys. Rev. Lett. {\bf 87}, 160406 (2001).

\bibitem{Hellweg-etal-02} D. Hellweg, S. Dettmer, P. Ryytty,
  J. J. Arlt, W. Ertmer, K. Sengstock, D. S. Petrov,
  G. V. Shlyapnikov, H. Kreutzmann, L. Santos, and M. Lewenstein,
  Phase Fluctuations in Bose-Einstein Condensates, Appl. Phys. B {\bf
    73}, 781 (2001).

\bibitem{Hellweg-etal-03} D. Hellweg, L. Cacciapuoti, M. Kottke,
  T. Schulte, K. Sengstock, W. Ertmer, and J. J. Arlt, Measurement of
  the Spatial Correlation Function of Phase Fluctuating Bose-Einstein
  Condensates, Phys. Rev. Lett. {\bf 91}, 010406 (2003).

\bibitem{Ritter-etal-07} S. Ritter, A. \"Ottl, T. Donner, T. Bourdel,
  M. K\"ohl, and T. Esslinger, Observing the Formation of Long-Range
  Order during Bose-Einstein Condensation, Phys. Rev. Lett. {\bf 98},
  090402 (2007).

\bibitem{BDZ-08} I. Bloch, J. Dalibard, and W. Zwerger, Many-body
  physics with ultracold gases, Rev.\ Mod.\ Phys.\ {\bf 80}, 885
  (2008).



\bibitem{DRBOKE-07} T. Donner, S. Ritter, T. Bourdel, A. \"Ottl,
  M. K\"ohl, and T. Esslinger, Critical behavior of a trapped
  interacting Bose gas, Science {\bf 315}, 1556 (2007).

\bibitem{DZZH-07} R.B. Diener, Q. Zhou, H. Zhai, and T.L. Ho, Criterion
  for Bosonic Superfluidity in an Optical Lattice,
  Phys. Rev. Lett. {\bf 98}, 180404 (2007).

\bibitem{BB-09} A. Bezett and P.B. Blakie, Critical properties of a
  trapped interacting Bose gas, Phys. Rev. A {\bf 79}, 033611 (2009).

\bibitem{CV-09} M. Campostrini and E. Vicari, Critical behavior and
  scaling in trapped systems, Phys. Rev. Lett. {\bf 102}, 240601
  (2009); (E) {\bf 103}, 269901 (2009); M. Campostrini and E. Vicari,
  Trap-size scaling in confined particle systems at quantum
  transitions, Phys. Rev. A {\bf 81}, 023606 (2010).



\bibitem{ZKKT-09} Q. Zhou, Y. Kato, N. Kawashima, and N. Trivedi, Direct
  Mapping of the Finite Temperature Phase Diagram of Strongly
  Correlated Quantum Models, Phys. Rev. Lett. {\bf 103}, 085701
  (2009).

\bibitem{Trotzky-etal-10} S. Trotzky, L. Pollet, F. Gerbier,
  U. Schnorrberger, I. Bloch, N.V. Prokofev, B. Svistunov, and
  M. Troyer, Suppression of the critical temperature for superfluidity
  near the Mott transition, Nat. Phys. {\bf 6}, 998 (2010).

\bibitem{HZ-10} T.-L. Ho and Q. Zhou, Obtaining the phase diagram and
  thermodynamic quantities of bulk systems from the densities of
  trapped gases, Nat. Phys. {\bf 6}, 131 (2010).

\bibitem{PPS-10} L. Pollet, N.V. Prokof'ev, and B.V. Svistunov, 
Criticality in Trapped Atomic Systems,
Phys. Rev. Lett. {\bf 104}, 245705 (2010).

\bibitem{NNCS-10}
S. Nascimbene, N. Nayon, F. Chevy, and C. Salomon,
The equation of state of ultracold Bose and Fermi gases: a few examples,
New J. Phys. {\bf 12}, 103026 (2010).

\bibitem{ZKKT-10} Q. Zhou, Y. Kato, N. Kawashima, and N. Trivedi, Direct
  Mapping of the Finite Temperature Phase Diagram of Strongly
  Correlated Quantum Models, Phys. Rev. Lett. {\bf 105}, 199601 (2010).

\bibitem{QSS-10} S.L.A.~de~Queiroz, R.R.~dos~Santos, and R.B.~Stinchcombe,
Finite-size scaling behavior in trapped systems,
Phys. Rev. E {\bf 81}, 051122 (2010).


\bibitem{FCMCW-11} S. Fang, C-M. Chung, P-N. Ma, P. Chen, and D-W. Wang,
Quantum criticality from in situ density imaging,
Phys. Rev. A {\bf 83}, 031605(R) (2011).


\bibitem{HM-11}
K. R. A. Hazzard and E. J. Mueller, 
Techniques to measure quantum criticality in cold atoms,
Phys. Rev. A {\bf 84}, 013604 (2011).

\bibitem{Pollet-12} L. Pollet, Recent developments in quantum Monte
  Carlo simulations with applications for cold gases,
  Rep. Prog. Phys. {\bf 75}, 094501 (2012).

\bibitem{CR-12} J. Carrasquilla and M. Rigol, Superfluid to normal
  phase transition in strongly correlated bosons in two and three
  dimensions, Phys. Rev. A {\bf 86}, 043629 (2012).


\bibitem{CTV-13} G. Ceccarelli, C. Torrero, and E. Vicari, Critical
  parameters from trap-size scaling in trapped particle systems,
  Phys. Rev. B {\bf 87} 024513 (2013).

\bibitem{CN-14} G. Ceccarelli and J. Nespolo, Universal scaling of
  three-dimensional bosonic gases in a trapping potential,
  Phys. Rev. B {\bf 89}, 054504 (2014).

\bibitem{CCBDWNDB-14}
L. Corman, L. Chomaz, T. Bienaim\'e, R. Desbuquois,
C. Wettenberg, S. Nascimbene, J. Dalibard, and J. Beugnon, 
Quench-induced supercurrents in an annular Bose gas,
Phys. Rev. Lett. {\bf 113}, 135302 (2014).

\bibitem{CNPV-15} G. Ceccarelli, J. Nespolo, A. Pelissetto, and
  E. Vicari, Bose-Einstein condensation and critical behavior of
  two-component bosonic gases, Phys. Rev. A {\bf 92}, 043613 (2015);
  Phase diagram and critical behaviors of mixtures of Bose gases,
  Phys. Rev. A {\bf 93}, 033647 (2016).


\bibitem{NGSH-15} N. Navon, A. L. Gaunt, R. P. Smith, and
  Z. Hadzibabic, Critical Dynamics of Spontaneous Symmetry Breaking in
  a Homogeneous Bose gas, Science {\bf 347}, 167 (2015).


\bibitem{CCBDWNBD-15} L. Chomaz, L. Corman, T. Bienaim\'e,
  R. Desbuquois, C. Wettenberg, S. Nascimbene, J. Beugnon, and
  J. Dalibard, Emergence of coherence via transverse condensation in a
  uniform quasi-two-dimensional Bose gas, Nat. Commun. {\bf 6}, 6162
  (2015).

\bibitem{DV-17} F. Delfino and E. Vicari, Critical behavior at the
  spatial boundary of a trapped inhomogeneous Bose-Einstein
  condensate, Phys. Rev. A {\bf 95}, 053606 (2017).


\bibitem{BN-17} J. Beugnon and N. Navon, Exploring the Kibble-Zurek
  mechanism with homogeneous Bose gases, J. Phys. B:
  At. Mol. Opt. Phys. {\bf 50}, 022002 (2017).

\bibitem{CDMV-16}
G. Ceccarelli, F. Delfino, M. Mesiti, and E. Vicari,
Shape dependence and anisotropic finite-size scaling of the
phase coherence of three-dimensional Bose-Einstein condensed gases,
Phys. Rev. A  {\bf 94}, 053609 (2016).



\bibitem{PSW-01} D. S. Petrov, G. V. Shlyapnikov, and
  J. T. M. Walraven, Phase-fluctuating 3D Bose-Einstein condensates in
  elongated traps, Phys. Rev. Lett. {\bf 87}, 050404 (2001).


\bibitem{Mathey-etal-10} 
L. Mathey, A. Ramanathan, K. C. Wright,
  S. R. Muniz, W. D. Phillips, and C. W. Clark, Phase fluctuations in
  anisotropic Bose-Einstein condensates: From cigars to rings,
  Phys. Rev. A {\bf 82}, 033607 (2010).

\bibitem{GCP-12} D. Gallucci, S. P. Cockburn, and N. P. Proukakis,
  Phase coherence in quasicondensate experiments: An {\em ab initio}
  analysis via the stochastic Gross-Pitaevskii equation,
Phys. Rev. A {\bf 86}, 013627 (2012).

\bibitem{RMHDTLK-13} W. RuGWay, A.G. Manning, S.S. Hodgman, R.G. Dall,
  A.G. Truscott, T. Lamberton, and K.V. Kheruntsyan, Observation of
  Transverse Bose-Einstein Condensation via Hanbury Brown-Twiss
  Correlations, Phys. Rev. Lett. {\bf 111}, 093601 (2013).

\bibitem{KT-73} J. M. Kosterlitz and D. J. Thouless, Ordering,
  metastability and phase transitions in two-dimensional systems,
  J.\ Phys. C: Solid State {\bf 6}, 1181 (1973)

\bibitem{B-72} V. L. Berezinskii, Destruction of Long-range Order in
  One-dimensional and Two-dimensional Systems having a Continuous
  Symmetry Group I. Classical Systems, Zh. Eksp. Theor. Fiz. {\bf 59},
  907 (1970) [Sov. Phys. JETP {\bf 32}, 493 (1971)].

\bibitem{Kosterlitz-74}
J. M. Kosterlitz, 
The critical properties of the two-dimensional xy model,
J. Phys. C {\bf 7}, 1046 (1974).

\bibitem{JKKN-77} J. V. Jos\'e, L. P. Kadanoff, S. Kirkpatrick, and
  D. R. Nelson, Renormalization, vortices, and symmetry-breaking
  perturbations in the two-dimensional planar model, Phys. Rev. B {\bf
    16}, 1217 (1977).

\bibitem{HKCBD-06} Z. Hadzibabic, P. Kr\"uger, M. Cheneau,
  B. Battelier, and J. Dalibard, Berezinskii-Kosterlitz-Thouless
  crossover in a trapped atomic gas, Nature {\bf 441}, 1118 (2006).

\bibitem{KHD-07}
P. Kr\"uger, Z. Hadzibabic, and J. Dalibard,
Critical Point of an Interacting Two-Dimensional Atomic Bose Gas,
Phys. Rev. Lett. {\bf 99}, 040402 (2007).

\bibitem{HKCRD-08} Z. Hadzibabic, P. Kr\"uger, M. Cheneau, S. P. Rath,
  and J. Dalibard, The trapped two-dimensional Bose gas: from
  Bose-Einstein condensation to Berezinskii-Kosterlitz-Thouless
  physics, New J. Phys. {\bf 10}, 045006 (2008).

\bibitem{CRRHP-09} P. Clad\'e, C. Ryu, A. Ramanathan, K. Helmerson,
  and W. D. Phillips, Observation of a 2D Bose Gas: From Thermal to
  Quasicondensate to Superfluid, Phys. Rev. Lett. {\bf 102}, 170401
  (2009).

\bibitem{HZGC-10}
C.-L. Hung, X. Zhang, N. Gemelke, and C. Chin,
Observation of scale invariance and universality in two-dimensional Bose gases,
Nature {\bf 470}, 236 (2011).

\bibitem{Pl-etal-11} T. Plisson, B. Allard, M. Holzmann, G. Salomon,
  A. Aspect, P. Bouyer, and T. Bourdel, Coherence properties of a
  two-dimensional trapped Bose gas around the superfluid transition,
  Phys. Rev. A {\bf 84}, 061606(R) (2011).

\bibitem{Desb-etal-12} R. Desbuquois, L. Chomaz, T. Ysefsah,
  J. L\'eonard, J. Beugnon, C. Weitenberg, and J. Dalibard, Superfluid
  behaviour of a two-dimensional Bose gas, Nat. Phys. {\bf 8}, 645
  (2012).

\bibitem{Barber-83}
M. N. Barber, Finite-size scaling in
{\em Phase Transitions and Critical Phenomena},
Vol. 8, eds. C. Domb abd J. L. Lebowitz
(Academic Press, 1983).

\bibitem{Privman-90}
{\em Finite Size Scaling and Numerical Simulations of Statistical Systems}, 
ed. V. Privman (World Scientific, 1990).


\bibitem{GKMD-08} F. M. Gasparini, M. O. Kimball, K. P. Mooney, and
  M. Diaz-Avilla, Finite-size scaling of $^4$He at the superfluid
  transition, Rev. Mod. Phys. {\bf 80}, 1009 (2008).


\bibitem{SM-95}
N. Schultka and E. Manousakis,
Crossover from two- to three-dimensional behavior in superfluids,
Phys. Rev. B {\bf 51}, 11712 (1995).

\bibitem{SM-96}
N. Schultka and E. Manousakis,
Scaling of superfluid density in superfluid films,
J. Low Temp. Phys. {\bf 105}, 3 (1996).

\bibitem{SM-97}
N. Schultka and E. Manousakis,
Boundary effects in superfluid films,
J. Low Temp. Phys. {\bf 109}, 733 (1997).

\bibitem{Hasenbusch-09} M. Hasenbusch, Kosterlitz-Thouless transition
  in thin films: A Monte Carlo study of three-dimensional lattice
  models, J. Stat. Mech.: Theory Expt.  P02005 (2009).


\bibitem{Kibble-76} T. W. B. Kibble, Topology of cosmic domains and
  strings, J. Phys. A {\bf 9}, 1387 (1976).

\bibitem{Zurek-85} W. H. Zurek, Cosmological experiments in superfluid
  helium?, Nature {\bf 317}, 505 (1985).


\bibitem{FWGF-89} M.P.A. Fisher, P.B. Weichman, G. Grinstein, and
  D.S. Fisher, Boson localization and the superfluid-insulator
  transition, Phys. Rev. B {\bf 40}, 546 (1989).

\bibitem{JBCGZ-98} D. Jaksch, C. Bruder, J.I. Cirac, C.W. Gardiner,
  and P. Zoller, Cold Bosonic Atoms in Optical Lattices,
  Phys. Rev. Lett. {\bf 81}, 3108 (1998).

\bibitem{CPS-07} B. Capogrosso-Sansone, N.V. Prokof'ev, and
  B.V. Svistunov, Phase diagram and thermodynamics of the
  three-dimensional Bose-Hubbard model, Phys. Rev. B {\bf 75}, 134302
  (2007).

\bibitem{PV-02}
A. Pelissetto and E. Vicari,
Critical Phenomena and Renormalization Group Theory,
Phys. Rep. {\bf 368}, 549  (2002).

\bibitem{Lipa-etal-96}
J.A. Lipa, D.R. Swanson, J.A. Nissen, T.C.P. Chui, 
and U.E. Israelsson, 
Heat Capacity and Thermal Relaxation of Bulk Helium very near the Lambda Point,
Phys. Rev. Lett. {\bf 76}, 944 (1996);
J.A.~Lipa, J.A.~Nissen, D.A.~Stricker, D.R.~Swanson, and T.C.P.~Chui,
Specific heat of liquid helium in zero gravity very near the lambda point,
Phys.  Rev.  B  {\bf 68}, 174518 (2003).

\bibitem{CHPV-06} M. Campostrini, M. Hasenbusch, A. Pelissetto, and
  E. Vicari, Theoretical estimates of the critical exponents of the
  superfluid transition in $^4$He by lattice methods, Phys. Rev. B
  {\bf 74}, 144506 (2006).

\bibitem{BMPS-06}
E. Burovski, J. Machta, N. Prokof'ev, and B. Svistunov,
High-precision measurement of the thermal exponent for the three-dimensional 
XY universality class,
Phys. Rev. B {\bf 74}, 132502 (2006).

\bibitem{GZ-98}
R. Guida and J. Zinn-Justin, 
Critical exponents of the N-vector model,
J. Phys. A {\bf 31}, 8103 (1998).

\bibitem{KPSV-16}
F. Kos, D. Poland, D. Simmons-Duffin, and A. Vichi
Precision Islands in the Ising and O($N$) Models,
JHEP {\bf 08} (2016) 036.


\bibitem{KP-17}
M. V. Kompaniets and E. Panzer,
Minimally subtracted six loop renormalization of 
O($n$)-symmetric $\varphi^4$ theory and critical exponents,
arXiv:1705.06483.

\bibitem{FBJ-73}
M. E. Fisher, M. N. Barber, and D. Jasnow, 
Helicity Modulus, Superfluidity, and Scaling in Isotropic Systems,
Phys. Rev. A {\bf 8}, 1111 (1973).

\bibitem{PC-87}
E. L. Pollock and D. M. Ceperley, 
Path-integral computation of superfluid densities,
Phys. Rev. B {\bf 36}, 8343 (1987).


\bibitem{MW-66} N.D. Mermin and H. Wagner, Absence of Ferromagnetism
  or Antiferromagnetism in One- or Two-Dimensional Isotropic
  Heisenberg Model, Phys. Rev. Lett. {\bf 17}, 1133 (1966).

\bibitem{H-67}
P. C. Hohenberg, 
Existence of Long-Range Order in One and Two Dimensions,
Phys. Rev. {\bf 158}, 383 (1967).


\bibitem{CNPV-13}
G. Ceccarelli, J. Nespolo, A. Pelissetto, and E. Vicari,
Universal behavior of two-dimensional bosonic gases at 
Berezinskii-Kosterlitz-Thouless transitions,
Phys. Rev.  B  {\bf 88}, 024517 (2013).


\bibitem{HK-97}
K. Harada and N. Kawashima,
Universal jump in the helicity modulus of the two-dimensional quantum XY model,
Phys. Rev. B {\bf 55}, R11949 (1997).

\bibitem{Ding-92}
H.-Q. Ding, 
Phase transition and thermodynamics of quantum XY model in two dimensions,
Phys. Rev. B {\bf 45}, 230 (1992).

\bibitem{DM-90}
H.-Q. Ding and M.S. Makivi\'c,
Kosterlitz-Thouless transition in the two-dimensional quantum XY model,
Phys. Rev. B {\bf 42}, 6827 (1990).


\bibitem{Fisher-71}
M. E. Fisher, {\em Critical Phenomena},
Proceedings of the International School of Physics Enrico Fermi,
edited by M. S. Green (Academic, New York, 1971).

\bibitem{CF-76}
T. W. Capehart and M. E. Fisher,
Susceptibility scaling functions for ferromagnetic Ising films,
Phys. Rev. B {\bf 13}, 5021 (1976).

\bibitem{It-Dr-book} C. Itzykson and J. M. Drouffe,
{\em Statistical Field Theory} (Cambridge Univ. Press, Cambridge, 1989).


\bibitem{CFT-book}
P. Di Francesco, P. Mathieu, and D. Senechal,
{\em Conformal Field Theory} (Springer Verlag, New York, 1997).


\bibitem{Hasenbusch-05} M. Hasenbusch, The two dimensional XY model at
  the transition temperature: a high precision numerical study,
  J. Phys. A {\bf 38}, 5869 (2005).

\bibitem{Gradstein} I. S. Gradshteyn and I. M. Ryzhik, 
{\em Table of  Integrals, Series, and Products}, 
edited by A. Jeffrey and D. Zwillinger, 7th edition 
(Academic Press, San Diego, 2007).



\bibitem{PV-13} A. Pelissetto and E. Vicari, Renormalization-group
  flow and asymptotic behaviors at the Berezinskii-Kosterlitz-Thouless
  transitions, Phys. Rev. E {\bf 87}, 032105 (2013).


\bibitem{HPV-05} M. Hasenbusch, A. Pelissetto, and E. Vicari,
  Multicritical behavior in the fully frustrated XY model and related
  systems, J. Stat. Mech.: Theory Expt.  P12002 (2005).


\bibitem{AGG-80} D. J. Amit, Y. Y. Goldschmidt, and G. Grinstein,
  Renormalisation group analysis of the phase transition in the 2D
  Coulomb gas, Sine-Gordon theory and XY-model, J. Phys. A {\bf 13},
  585 (1980).

\bibitem{HMP-94}
M. Hasenbusch, M. Marcu, and K. Pinn,
High precision renormalization group study of the roughening transition,
Physica A {\bf 208}, 124 (1994).

\bibitem{HP-97} M. Hasenbusch and K. Pinn, Computing the roughening
  transition of Ising and solid-on-solid models by BCSOS model
  matching, J. Phys. A {\bf 30}, 63 (1997).

\bibitem{Balog-01}
J. Balog, 
Kosterlitz-Thouless theory and lattice artifacts,
J. Phys. A {\bf 34}, 5237 (2001).

\bibitem{Hasenbusch-08}
M. Hasenbusch,  
The Binder cumulant at the Kosterlitz-Thouless transition,
J. Stat. Mech.: Theory Expt. P08003 (2008).


\bibitem{Hasenbusch-12} 
M. Hasenbusch, 
Thermodynamic Casimir effect: Universality and corrections to scaling,
Phys. Rev. B {\bf 85}, 174421 (2012).

\bibitem{Binder-87} K. Binder, Theory of first-order phase
  transitions, Rep. Prog. Phys. {\bf 50}, 783 (1987).

\bibitem{Bray-94} A.J. Bray, 
Theory of phase-ordering kinetics,
Adv. Phys. {\bf 43}, 357 (1994).

\bibitem{CG-05} P. Calabrese and A. Gambassi, Ageing Properties of
  Critical Systems, J. Phys. A {\bf 38}, R133 (2005).

\bibitem{GZHF-10} S. Gong, F. Zhong, X. Huang, and S. Fan, Finite-time
  scaling via linear driving, New J. Phys. {\bf 12}, 043036 (2010).

\bibitem{PSSV-11} A. Polkovnikov, K. Sengupta, A. Silva, and
  M. Vengalattore, Colloquium: Nonequilibrium dynamics of closed
  interacting quantum systems, Rev. Mod. Phys. {\bf 83}, 863 (2011).

\bibitem{CEGS-12} A. Chandran, A. Erez, S. S. Gubser, and S. L. Sondhi,
  Kibble-Zurek problem: Universality and the scaling limit,
  Phys. Rev. B {\bf 86}, 064304 (2012).

\bibitem{Braun-etal-15} S. Braun, M. Friesdorf, S.S. Hodgman,
  M. Schreiber, J.P. Ronzheimer, A. Riera, M. del Rey, I. Bloch,
  J. Eisert, and U. Schneider, Emergence of coherence and the dynamics
  of quantum phase transitions, PNAS {\bf 112}, 3641 (2015).

\bibitem{Biroli-15} G. Biroli, Slow Relaxations and Non-Equilibrium
  Dynamics in Classical and Quantum Systems, arXiv:1507.05858.

\bibitem{PV-16} A. Pelissetto and E. Vicari, Off-equilibrium scaling
  behaviors driven by time-dependent external fields in
  three-dimensional O($N$) vector models, Phys. Rev. E {\bf 93},
  032141 (2016).

\bibitem{DWGGP-16} M. J. Davis, T. M. Wright, T. Gasenzer,
  S. A. Gardiner, and N. P. Proukakis, Formation of Bose-Einstein
  condensates, arXiv:1601.06197.

\bibitem{ARBBHC-16}
M. Anquez, B.A. Robbins, H.M. Bharath, M.J. Boguslawski, T.M. Hoang, 
and M.S. Chapman,
Kibble-Zurek Mechanism in a Spin-1 Bose-Einstein Condensate,
Phys. Rev. Lett. {\bf 116}, 155301 (2016).

\bibitem{PV-17} A. Pelissetto and E. Vicari, Dynamic off-equilibrium
  transition in systems slowly driven across thermal first-order
  transitions, Phys. Rev. Lett. {\bf 118}, 030602 (2017).

\bibitem{HH-77}
P. C. Hohenberg and B. I. Halperin,
Theory of dynamic critical phenomena,
Rev. Mod. Phys. {\bf 49}, 435 (1977).

\bibitem{FM-06}
R. Folk and G. Moser, 
Critical dynamics: A field-theoretical approach, 
J. Phys. A {\bf 39}, R207 (2006).


\bibitem{SHLVS-06}
L. E. Sadler, J. M. Higbie, S. R. Leslie, M. Vengalattore,
and D. M. StamperKurn, Spontaneous symmetry breaking in a quenched
ferromagnetic spinor Bose-Einstein consensate,
Nature {\bf 443}, 312 (2006).

\bibitem{WNSBD-08}
C. N. Weiler, T. W. Neely, D. R. Scherer, A. S. Bradley,
M. J. Davis, and B. P. Anderson,
Spontaneous vortices in the formation of Bose-Eistein 
condensates, Nature {\bf 455}, 948 (2008).



\bibitem{LDSDF-13} G. Lamporesi, S. Donadello, S. Serafini,
  F. Dalfovo, and G. Ferrari, Spontaneous creation of Kibble–-Zurek
  solitons in a Bose-–Einstein condensate, Nat. Phys. {\bf 9}, 656
  (2013).


\bibitem{vDK-97}
N.J. van Druten and W. Ketterle,
Two-step condensation of the ideal Bose gas in highly anisotropic traps,
Phys. Rev. Lett. {\bf 79}, 549 (1997).

\bibitem{AJKB-11}
J. Armjio, T. Jacqmin, K. Kheruntsyan, and I. Bouchoule,
Mapping out the quasicondensate transition through the dimensional
crossover from one to three dimensions,
Phys. Rev. A {\bf 83}, 021605 (2011).


\bibitem{SK-91} A. W. Sandvik and J. Kurlij\"arvi, Quantum Monte Carlo
  simulation method for spin systems, Phys. Rev. B {\bf 43}, 5950
  (1991).

\bibitem{SS-02} O. F. Sylju\aa{}sen and A. W. Sandvik, Quantum Monte
  Carlo with directed loops, Phys. Rev. E {\bf 66}, 046701 (2002).

\bibitem{DT-01} A. Dorneich and M. Troyer, Accessing the dynamics of
  large many-particle systems using the stochastic series expansion,
  Phys. Rev. E {\bf 64}, 066701 (2001).


\end{thebibliography}
\end{document}